\documentclass[twocolumn]{aastex631}
\usepackage{csquotes}
\usepackage{amsmath}
\usepackage{mathtools}
\usepackage[inline]{enumitem} 
\usepackage{float}
\usepackage{ulem}
\usepackage{xcolor}
\DeclareRobustCommand{\erase}{\bgroup\markoverwith{\textcolor{black}{\rule[.5ex]{2pt}{0.4pt}}}\ULon}

\graphicspath{{./}{figures/}}
\usepackage{filecontents}

\usepackage{hyperref}

\usepackage{xcolor}

\usepackage{float}
\usepackage{graphicx}
\usepackage{booktabs}

\shortauthors{takahashi et al.}

\newcommand{\emailaddress}{satoko.takahashi@nao.ac.jp}

\begin{document}

\title{An Extremely Young Protostellar Core, MMS\,1/ OMC-3: 
Episodic Mass Ejection History Traced by the Micro SiO Jet}

\correspondingauthor{Satoko Takahashi}

\author[0000-0002-7287-4343]{Satoko Takahashi}
\affiliation{\textnormal{National Astronomical Observatory of Japan, 2-21-1 Osawa, Mitaka, Tokyo 181-8588, Japan; \href{mailto:\emailaddress}{\emailaddress}}}
\affiliation{\textnormal{Astronomical Science Program, The Graduate University for Advanced Studies, SOKENDAI, 2-21-1 Osawa, Mitaka, Tokyo 181-8588, Japan}}

\author[0000-0002-0963-0872]{Masahiro N. Machida}
\affiliation{\textnormal{Department of Earth and Planetary Sciences, Faculty of Science, Kyushu University, 744 Motooka, Nishi-ku, Fukuoka 819-0395, Japan}}

\author[0000-0002-7951-1641]{Mitsuki Omura}
\affiliation{\textnormal{Department of Earth and Planetary Sciences, Graduate School of Science, Kyushu University, 744 Motooka, Nishi-ku, Fukuoka 819-0395, Japan}}

\author[0000-0002-6773-459X]{Doug Johnstone}
\affiliation{\textnormal{NRC Herzberg Astronomy and Astrophysics, 5071 West Saanich Rd, Victoria, BC, V9E 2E7, Canada }}
\affiliation{\textnormal{Department of Physics and Astronomy, University of Victoria, Victoria, BC, V8P 5C2, Canada}}

\author[0000-0003-1549-6435]{Kazuya Saigo}
\affiliation{\textnormal{Department of Physics and Astronomy, Graduate School of Science and Engineering, Kagoshima University, 1-21-35 Korimoto,
Kagoshima,Kagoshima 890-0065, Japan}}

\author[0000-0002-8217-7509]{Naoto Harada}
\affiliation{\textnormal{Department of Earth and Planetary Sciences, Graduate School of Science, Kyushu University, 744 Motooka, Nishi-ku, Fukuoka 819-0395, Japan}}

\author[0000-0003-2726-0892]{Kohji Tomisaka}
\affiliation{\textnormal{National Astronomical Observatory of Japan, 2-21-1 Osawa, Mitaka, Tokyo 181-8588, Japan; \href{mailto:\emailaddress}{\emailaddress}}}

\author[0000-0002-3412-4306]{Paul T. P. Ho}
\affiliation{\textnormal{Institute of Astronomy and Astrophysics, Academia Sinica, 11F of Astronomy-Mathematics Building, AS/NTU No. 1, Sec. 4, Roosevelt
Rd, Taipei 10617, Taiwan, R.O.C.}}
\affiliation{\textnormal{East Asian Observatory, 660 N. A’ohoku Place, Hilo, HI 96720, USA}}

\author[0000-0003-2343-7937]{Luis A. Zapata}
\affiliation{\textnormal{Instituto de Radioastronomia y Astrof\'{i}sica, Universidad Nacional Aut\'onoma de M\'exico, PO Box 3-72, 58090 Morelia, Michoac\' an, M\'exico}}

\author[0000-0002-6956-0730]{Steve Mairs}
\affiliation{\textnormal{East Asian Observatory, 660 N. A’ohoku Place, Hilo, HI 96720, USA}}
\affiliation{\textnormal{NRC Herzberg Astronomy and Astrophysics, 5071 West Saanich Rd, Victoria, BC, V9E 2E7, Canada }}

\author[0000-0002-7154-6065]{Gregory J. Herczeg}
\affiliation{\textnormal{Kavli Institute for Astronomy \& Astrophysics, Peking University, Yiheyuan Lu 5, Haidian Qu, 100871 Beijing, People's Republic of China }}
\affiliation{\textnormal{Department of Astronomy, Peking University, Yiheyuan 5, Haidian Qu, 100871 Beijing, People's Republic of China}}

\author[0000-0003-2726-0892]{Kotomi Taniguchi}
\affiliation{\textnormal{National Astronomical Observatory of Japan, 2-21-1 Osawa, Mitaka, Tokyo 181-8588, Japan; \href{mailto:\emailaddress}{\emailaddress}}}

\author[0009-0009-2263-5502]{Yuhua Liu}
\affiliation{\textnormal{Department of Earth and Planetary Sciences, Graduate School of Science, Kyushu University, 744 Motooka, Nishi-ku, Fukuoka 819-0395, Japan}}

\author[0000-0001-5817-6250]{Asako Sato}
\affiliation{\textnormal{Department of Earth and Planetary Sciences, Graduate School of Science, Kyushu University, 744 Motooka, Nishi-ku, Fukuoka 819-0395, Japan}}



\begin{abstract}

We present ${\sim}0.2$\,arcsec ($\sim$80\,au) 
resolution observations of the CO\,(2--1) 
and SiO\,(5--4) lines made with the Atacama 
large millimeter/submillimeter array toward an extremely 
young intermediate-mass protostellar source (t$_{\rm dyn}<$1000\,years), 
MMS\,1 located in the Orion Molecular Cloud-3 region. 
We have successfully imaged a very compact 
CO molecular outflow associated with MMS\,1, 
having deprojected lobe sizes of $\sim$18000\,au 
(red-shifted lobe) and $\sim$35000\,au 
(blue-shifted lobe).  
We have also detected an extremely compact 
($\lesssim$1000\,au) and collimated SiO 
protostellar jet within the CO outflow. 
The maximum deprojected jet speed is measured to be 
as high as 93\,km\,s$^{-1}$. 
The SiO jet wiggles and displays a chain of knots. 
Our detection of the molecular outflow and jet is the first direct evidence that 
MMS\,1 already hosts a protostar. 
The position-velocity diagram obtained 
from the SiO emission shows two distinct structures: 
(i) bow-shocks associated with 
the tips of the outflow, and 
(ii) a collimated jet, showing the jet velocities 
linearly increasing with the distance 
from the driving source. 
Comparisons between the observations 
and numerical simulations quantitatively 
share similarities such as 
multiple-mass ejection events within the jet 
and Hubble-like flow associated 
with each mass ejection event. 
Finally, while there is a weak flux decline seen in the 850\,$\mu$m light curve obtained with JCMT/SCUBA\,2 toward MMS\,1, no dramatic flux change events are detected. This 
suggests that there has not been a clear burst event within the last 8 years.


\end{abstract}


\keywords{Star formation (1569) --- Stellar jets (1807) --- 
Stellar winds (1636), Protostars (1302) --- Shocks (2086)} 


\section{Introduction}\label{sec:intro}

Molecular outflows and jets are observed in low- to high-mass 
star-forming regions (\citealt{bachiller1999, arce2007,bally2016, lee2020} 
and references therein). 
They are spatially extended, hence a useful tool to explore 
protostars when they are still deeply embedded within the core, 
and even when infrared counterparts are not detected 
at the peak position of the millimeter source 
(e.g., \citealt{takahashi2009, takahashi2012b, takahashi2019}). 
Molecular outflows are ubiquitously observed toward protostars. 
Observationally, the molecular outflows are accompanied with 
a wide-opening angle cavity (or also called the outflow lobe), 
having the low- and intermediate-velocity ranges ($\lesssim$50\,km\,s$^{-1}$). 
Protostellar jets are known as 
the extremely high velocity flow (a.k.a. EHV flow; \citealt{bachiller1990}), 
which shows a collimated morphology with high velocity gas 
($\gtrsim$50\,km\,s$^{-1}$).     
Whereas the molecular outflows are ubiquitously observed, 
only seven EHV flows had been 
reported before ALMA.
Furthermore, detection of the EHV jets was limited to bright Class 0 sources (\citealt{bachiller1990, bachiller1991a, bachiller1991b, bachiller1996, lebron2006, hirano2010, gomez2013, tafalla2017, lee2017, lee2020}).
This field of study has grown with the improved sensitivity of ALMA, which enables us to image more EHV flows associated with both Class~0 and Class~I protostellar stages  (e.g., \citealt{matsushita2019,Jhan2022,Dutta2022, Dutta2023}).

In addition to sensitivity improvements, 
ALMA sub-arcsecond observations 
enable us to detect not only bright outflows and jets in the sky, 
but also fainter or compact (i.e., younger) ones 
(e.g., \citealt{zapata2015, matsushita2019, tokuda2020, morii2021}). 
Searching for very compact molecular outflows and jets 
($\lesssim$ a few 1000\,au) sheds light on 
the first stage of the star formation process 
such as formation of the first adiabatic cores 
(e.g., \citealt{larson1969, masunaga2000, saigo2006,maureira2020}) 
and protostar formation immediately following the second collapse 
(e.g., \citealt{machida07, 
takahashi2012b, Takahashi2013b, Hirano2014, Gerin2015}). 

The outflow and jet are driven 
by magnetocentrifugal and magnetic pressure gradient mechanisms 
(\citealt{blandford1982, uchida1985, lynden-Bell2003}). 
ALMA's high angular resolution data 
can directly image the region near the launching point 
(\citealt{Bjerkeli2016}), 
allowing for a direct comparison with molecular outflows 
and jets produced by magnetohydrodynamics (MHD) models 
(e.g., \citealt{Shu1994, Tomisaka2002, machida07, Shang2023}). 
With this in mind, we can push 
in the right direction to understand 
better the driving mechanisms of the jets and outflows.

Recent studies present evidence for 
time variable mass ejection and accretion phenomena more commonly \citep[see review by][]{Fischer2023}. 
Periodically located knots (i.e., chain of knots) 
are detected by ALMA within some of the outflows and jets, using CO and SiO emission.
These knots may be related to a periodic variation in the jet velocity 
and periodic change in the mass ejections (\citealt{plunkett2015,matsushita2019, Jhan2022, Dutta2022, Dutta2023}). 
Dynamical timescales of each knot range 
from a few years to several thousands years. 
Non-steady mass accretion phenomena have also been suggested 
toward some of the protostellar sources through chemical 
diagnostics (e.g., \citealt{jorgensen2013, sharma2020}). 
The JCMT Transient Survey 
(\citealt{Herczeg2017}) has been monitoring 
$\sim$295 submillimeter bright sources within the Gould Belt, finding 18 protostellar 
sources showing secular variabilities after four years 
of their survey (\citealt{YHLee2021}). 
Furthermore, analysis of mid-infrared photometric monitoring observations 
by NEOWISE revealed that approximately 55\% of protostars 
show variabilities in their flux with a timescale of within a day to years (\citealt{Park2021}). 
The fraction of variable sources is higher 
at the earlier evolutionary stages than at the later evolutionary stages, and the sources show either long-term secular variability 
(linear, curved, and periodic) or short-term stochastic variability (burst, drop, and irregular). 
Such variations are expected to be related to 
activities around the stellar surface and inner disk edge, 
which likely affect the mass accretion rate onto the star. 
\cite{Zakri2022} suggested that bursts from Class 0 protostars are as frequent, or even more frequent than those from the Class I protostars based on a long-term ($\gtrsim$15~yr) systematic photometry study using Spitzer, WISE, and NEOWISE. 
Long term optical emission studies ($\sim$10~yr) toward T Tauri sources give us some hints of a possible link between the magnetospheric accretion and jet knots (e.g., time scale associated with 2-6 yr; \citealt{Takami2020, Takami2023}). 
All these studies demonstrate the importance of 
non-steady phenomena associated with the star-forming activities. 

The millimeter source MMS\,1, driving the outflows and jets 
presented in this paper, is located in 
the Orion Molecular Cloud\,-3 region 
(OMC-3 region; $d$=393\,pc by \citealt{Tobin2020}\footnote{The distance to the Orion A molecular 
cloud region is measured in several independent 
experiments: 414${\pm}$7\,pc \citep{menten2007} 
and 389$^{+24}_{-21}$\,pc \citep{sandstrom2007} 
using non-thermal radio emission 
in the Orion Nebula, 
437${\pm}$19\,pc \citep{hirota2007} 
and 416${\pm}$6\,pc \citep{kim2008} 
using masers toward Orion\,KL; 
398${\pm}$7\,pc using centimeter radio source 
toward Orion\,A molecular complex \citep{kounkel2017}, 
and 389${\pm}$3\,pc using APOGEE-2 
and $Gaia$ DR2 \citep{kounkel2018} 
and 393\,pc \citep{Tobin2020} using 
$Gaia$ DR2 toward Orion\,A complex. 
There are no direct distance 
measurements toward MMS\,1/OMC-3, 
we therefore rely on the nearby sources that 
have reliable parallax measurements. 
In this paper, \citet{Tobin2020} was adopted. 
All the measurements are consistent within $\sim$10\% of the adopted value.}). 
The region is known as one of the most active 
and young nearby intermediate-mass star-forming regions. 
There are 10 bright millimeter sources in the OMC-3 region 
(\citealt{chini1997,lis1998}) where several 
Class\,0 and Class\,I sources were 
identified from the previous multi-wavelength 
and multi-line observations 
(e.g., \citealt{chini1997, yu1997, lis1998,Johnstone1999, 
aso2000,stanke2002, williams2003, takahashi2006,
takahashi2008, takahashi2009, takahashi2012a, 
takahashi2013, takahashi2012b, megeath2012, stutz2013, 
furlan2016, Tobin2020} and references therein).

MMS\,1 was first identified by the IRAM 30m telescope 
(${\theta}{\sim}11\arcsec$) in the 1.3\,mm continuum band \citep{chini1997}. 
MMS\,1 was also identified as CSO\,5 
based on observations of SHARC/CSO in the 350\,$\mu$m 
continuum band \citep{lis1998}. 
Later the source was observed in the 850\,$\mu$m 
continuum emission using the Submillimeter array 
(${\theta}{\sim}4$\arcsec$.5$) 
and named as SMM\,2 \citep{takahashi2013}. 
MMS 1 is similarly bright as other 
(sub)millimeter sources in the OMC-3 region, 
which already host known protostars \citep{chini1997,lis1998,takahashi2013}.
The flux concentration ratio of $\sim$0.3 measured 
by \cite{takahashi2013}, which compares fluxes 
measured from the JCMT 14$\arcsec$ and SMA 4$\arcsec$.5 beams, 
is comparable with these other protostellar sources, 
indicating that a similar amount of material has
already condensed at the center of the core. 
Nevertheless, the source was not identified 
in the Herschel Orion Protostar Survey (HOPS survey) 
since no bright near- and mid- infrared compact 
and isolated emission was detected toward MMS\,1 
at wavelengths $\lesssim$70\,$\mu$m 
(\citealt{megeath2012, furlan2016}). 
No centimeter jet nor outflow has been reported toward MMS\,1 
as yet (\citealt{reipurth1999, takahashi2008}). 
Based on these facts, 
MMS\,1 has been considered to be either at the end of the prestellar phase, 
the first core phase, or the earliest  protostellar phase. 

With the improved angular resolution 
and sensitivity achieved with ALMA, 
for the first time, 
we have detected a very compact outflow and jet 
associated with MMS\,1 in both SiO\,(5--4) and CO\,(2--1) emission, respectively. 
We conclude that MMS 1 is in the extremely early 
evolutionary Class 0 stage 
even without association of a bright mid-infrared source. 
Studying the velocity structure of the jet, 
we find evidence of intermittent mass 
ejection within the MMS\,1 jet. 
We present these results in this paper. 

This paper is organized as follows. 
The observations are described in Section~\ref{sec:obs}. 
The results including the 1.3 mm continuum, CO\,(2--1), 
and SiO\,(5--4) data sets are described in Section~\ref{sec:results}. 
The velocity structure of the outflow and jet, 
and comparison with magnethydrodynamic (MHD) simulations 
are described in Section~\ref{sec:discussion}. 
Finally, concluding remarks and future prospects are given in Section~\ref{sec:sum}.

\section{OBSERVATIONS AND DATA REDUCTION} \label{sec:obs}
The ALMA observations were made in Cycle\,3, 
2016 January 29 (low angular resolution) 
and 2016 September 18 and 19 (high angular resolution) 
using the 1.3\,mm band (Band 6). 
The phase center was set toward the location of 
the millimeter source, MMS\,1: 
R.A. (J2000) = 5$^h$35$^m$18$^s$.03, 
decl. (J2000) =${-}$05$^{\circ}$00$'$17$\arcsec$.770. 
The observing parameters are listed 
in Table~\ref{table:obsparms}.  
Total on-source time is about 16 minutes 
with $\sim$40 antennas (high-angular resolution) 
and 4 minutes with 48 antennas (low-angular resolution). 
The low- and high-angular resolution data sets 
cover projected baselines between 
8.5\,k$\lambda$ and 239\,k$\lambda$ and 
between 8.9\,k$\lambda$ and 2416\,k$\lambda$, 
respectively. 
According to the ALMA proposer's guide, 
the maximum recoverable scale (MRS) of the low- 
and high-angular resolution data is approximately 
${\sim}9$\arcsec$.8$ and ${\sim}2$\arcsec$.0$, respectively.
We observed CO\,($J$=2--1) and SiO\,$(J$=5--4) 
at the velocity resolutions of 0.37\,km\,s$^{-1}$ 
and 0.39\,km\,s$^{-1}$ per channel, respectively. 
Line free channels corresponding to 
the effective band width of 454\,MHz 
are allocated for imaging the continuum data. 
The Common Astronomy Software Application 
(CASA; \citealt{CASA2022}) version 4.6.0 
and version 4.7.0 were used as the standard 
ALMA data reduction for the low- 
and high-angular resolution data sets, respectively. 

After calibration, the CLEANed images were 
made using the CASA task ``\texttt{tclean}''. 
The Briggs weighting with robust parameter 
of 0.5 were used for the final images. 
The velocity width of 5.0\,km\,s$^{-1}$ 
was used in order to produce the CO and SiO image cubes 
both for high- and low-angular resolution data sets. 
The resulting synthesized beam sizes and 1$\sigma$ 
rms noise levels for the CO, SiO, and 
the 1.3\,mm continuum emission are listed 
in Table~\ref{table:obsparms}.

\begin{table*}[ht!]
{\scriptsize
\begin{center}
\caption{\small Observing Parameters}
\label{table:obsparms}
\begin{tabular}{lcc}
\hline\hline \noalign {\smallskip}
Parameter & High-resolution Data & Low-resolution Data \\
\hline
Observing date (YYYY-MM-DD) & 2016-09-18 and -19    &   2016-01-29  \\
Number of antennas          & 38, 40		&  48	         \\
Primary beam size (arcsec)  & 27                    &   27		      \\
PWV (mm)     	  			& 1.1 -- 2.2            &   $\sim$2.6      \\
Phase stability rms (degree)\tablenotemark{$^a$} 		&	21 -- 52	     &	   $\sim$13	   	     \\
Bandpass calibrators        & J0510+1800            & J0522-3627  \\
Flux calibrators            & J0510+1800,J0522-3627 & J0522-3627  \\
Phase calibrators\tablenotemark{$^b$}   & J0607-0834       & J0541-0541        \\
Spectral line setups USB/LSB (GHz)   & \multicolumn{2}{c}{230.535; 231.319 / 217.102; 219.557} \\
Total continuum bandwidth; USB+LSB (MHz)    & 454           & 454            \\
Projected baseline ranges (k$\lambda$)  & 8.9 -- 2416	    & 8.5 -- 239      \\
Maximum recoverable size (arcsec)    & $\sim$2.0   & $\sim$9.8   \\
Total on-source time (minutes)       & 16     	 	& 4     \\       
Synthesized beam size of the CO (2--1) images (arcsec)  & 0.18$\times$0.15 (p.a.=-18deg.)   & 1.5$\times$0.9 (p.a.=-78deg.) \\
Synthesized beam size of the SiO (5--4) images (arcsec) & 0.21$\times$0.17 (p.a.=-27deg.)   & 1.6$\times$0.9 (p.a.=-78deg.)  \\
Synthesized beam size of the continuum images (arcsec) & 0.19$\times$0.16 (p.a.=-25deg.)    & 1.6$\times$0.9 (p.a.=-77deg.)   \\
RMS noise level of the CO (2--1) images (mJy beam$^{-1}$ km s$^{-1}$)\tablenotemark{$^c$}	    & 2.7	  & 3.8  \\
RMS noise level of the SiO (5--4) images (mJy beam$^{-1}$ km s$^{-1}$)\tablenotemark{$^c$}		& 1.8     & 3.4 \\
RMS noise level of the continuum images (mJy beam$^{-1}$) & 0.24         & 0.55  \\
\hline \noalign {\smallskip}
\end{tabular}
\end{center}}
\footnotesize $^a${Antenna-based phase differences on phase calibrators.}\\
$^b${The phase calibrator was observed every 8 minutes.}\\
$^c${RMS noise levels measured with the velocity width of 5 km s$^{-1}$.}\\
\end{table*}

\section{RESULTS} \label{sec:results}

We present the 1.3\,mm continuum, CO\,(2--1), and SiO\,(5--4) 
emission maps obtained with the ALMA observations in this section. 
In this paper, the terminologies of ``outflow'' 
and ``jet'' use the following definitions. 
The outflow shows as relatively low velocity gas 
($|v_{\rm{LSR}}-v_{\rm{sys}}|{\lesssim}$50\,km s$^{-1}$), 
and wide opening angles. 
The jet shows as high velocity outflow gas 
($|v_{\rm{LSR}}-v_{\rm{sys}}|{\gtrsim}$50\,km s$^{-1}$), 
and a collimated structure. 
The jets are located within the outflow lobes. 
The systemic velocity of 11\,km\,s$^{-1}$ is adopted for the system, 
which is determined from the optically thin molecular lines in the OMC-3 region 
(e.g., \citealt{takahashi2009, matsushita2019, morii2021}).

\subsection{The 1.3\,mm Continuum Emission} \label{subsec:cont}

\begin{figure*}
\epsscale{1.0}
\plotone{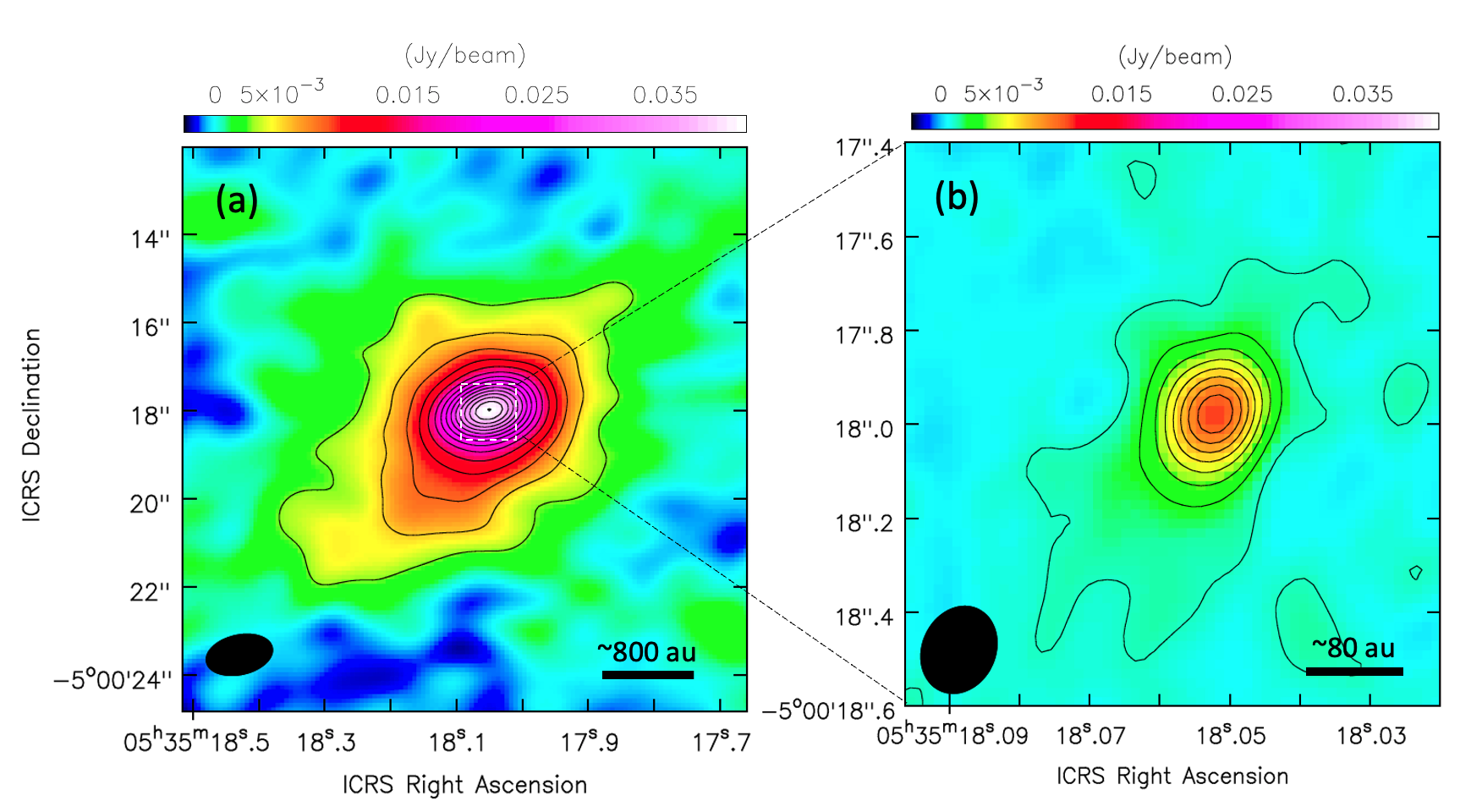}
\caption{The 1.3 mm continuum emission 
toward MMS 1 obtained from ALMA observations with
low angular resolution (a) and  high angular resolution (b). Contour level starts from 5$\sigma$ with an interval of 5$\sigma$ up to 70$\sigma$ for panel (a) and up to 40$\sigma$ for panel (b).}
\label{fig:f1}
\end{figure*}

Figure~\ref{fig:f1} shows the 1.3\,mm continuum emission 
toward MMS\,1 obtained with ALMA.
We have detected centrally concentrated continuum 
emission both in the low- and high-angular resolution images. 
Note that the origin of the 1.3\,mm continuum emission 
observed toward the protostellar sources 
in OMC-3 is considered to be significantly 
dominated by the thermal dust emission. The emission 
attributed to the free-free emission, tracing ionized jets, 
is likely negligible as estimated by \citet{takahashi2006,takahashi2009,takahashi2013,takahashi2019}. 

In order to characterize the source structure, 
particularly the centrally concentrated emission, 
presumably tracing the dust disk, 
we performed two-component two dimensional (2D) 
Gaussian fitting on the high-angular resolution image. 
The residual level of the fitting result is less than 5.6\% with respect to the observed peak flux 
(i.e., the residual level is less than S/N = 2.5).
The source structure was fitted by 
an extended (${\sim}750{\pm}47\,{\rm au}{\times}510{\pm}33$\,au) and 
a compact (${\sim}43{\pm}5.1\,{\rm au}{\times}26{\pm}4.7$\,au) Gaussian component, both of which are listed in Table~\ref{table:2Dgauss}. 
The position angle\footnote{The position angle is measured 
from the north to the east with respect 
to the blue-shifted lobe.} 
(p.a.) of the major axis of both extended 
and compact components 
($150^{\circ}\pm7.5^{\circ}$ 
and $156^{\circ}\pm12^{\circ}$, respectively) 
are roughly aligned to the direction 
of the large-scale filament in this region 
(i.e., \citealt{chini1997,Johnstone1999}; p.a.${\sim}135^{\circ}$). 
The orientation of the compact component is almost perpendicular to 
the molecular jet detected with ALMA 
in the SiO emission (Section~\ref{subsec:outflowjet}). 
Assuming that the detected compact 1.3\,mm continuum emission 
traces the dust disk, an inclination angle of the disk ($i$) 
is estimated to be ${\sim}53^{\circ}$ using 
the axis ratio of the 1.3\,mm continuum compact component. 
Note that the millimeter continuum fitting 
was also performed by \cite{Liu2023} 
as a part of the full polarization data analysis. 
The data were obtained in the 1.1\,mm band. 
A slightly higher angular resolution than that presented here
of $0$\arcsec$.14$ was achieved with 
a factor of 6.8 better sensitivity for 
the full polarization data presented. 
Despite the significant sensitivity differences, 
the total flux measured within the compact component 
is consistent within the factor of $\sim$1.2. 
The measured size of the disk-like structure 
in this paper is factor of 
$\sim$4.5 larger (in terms of the surface area) than 
those estimated in \cite{Liu2023}. 
This is likely due to the factor of $\sim$1.7 
larger beam size (in terms of the beam surface area) 
in the data set presented here compared 
with that presented in \cite{Liu2023} to ApJ).

The mass of the circumstellar material can be estimated 
from the 1.3\,mm continuum emission using the following equation, 
\begin{equation}
M_{\rm{H_2}} = \frac{F_{\lambda}d^2}{{\kappa_{\lambda}}B_{\lambda}(T_{d})},
\end{equation}
where $F_{\lambda}$ is the total 1.3\,mm flux, 
$d$ is the distance to the source, ${\kappa}_{\lambda}$ 
is the dust opacity (absorption coefficient per unit gas mass). 
Here, a gas-to-dust mass ratio of 100 is assumed. 
$T_d$ is the dust temperature, and $B_{\lambda}(T_d)$ 
is the Planck function at a temperature of $T_d$. 
Assuming, $d$ = 393\,pc 
(distance to the OMC-3 region measured from 
the $Gaia$ DR2 data; \citealt{Tobin2020, McBride2019, Gaia2018}), 
${\kappa}_{\rm{1.3mm}}$ = 0.011\,cm$^2$ g$^{-1}$ from 
the dust coagulation model of the MRN \citep{mathis1977} 
with thin ice mantles at a number density of 10$^{8}$\,cm$^{-3}$ 
\citep{ossenkopf1994}, $T_{\rm{dust}}$ = 30\,K -- 100\,K 
(e.g., \citealt{nomura2005}; \citealt{tomida2017}), and 
given the measured total fluxes listed in Table~\ref{table:2Dgauss}, 
the masses of the compact and extended components are estimated 
to be $M_{\rm{H_2(comp.)}}{\sim}${0.0051}--0.020\,$M_{\odot}$ 
and $M_{\rm{H_2(ext.)}}{\sim}$0.041--{0.16}\,$M_{\odot}$, respectively.
Note that the dust temperature assumption was discussed 
by \cite{Liu2023} in detail based on 
recent studies toward Orion protostars by \cite{Tobin2020} and \cite{xu2021} 
considering the stellar radiation and disk accretion heating, respectively. 
The minimum and maximum temperatures adopted here are 
comparable to the temperature assuming the stellar radiation 
expected from the given radius corresponding to 
the extended and compact structures. 

\begin{table*}
{\scriptsize
\begin{center}
\caption{\small Two Dimensional Gaussian Fitting Results obtained from the 1.3 mm Continuum Emission}
\label{table:2Dgauss}
\begin{tabular}{lccccc}
\hline\hline \noalign {\smallskip}
 & R.A. & Decl. & Deconvolved Size, p.a. & Peak Intensity & Flux Density \\
 & (J2000) & (J2000) & (arcsec, degree) & (mJy beam$^{-1}$) & (mJy) \\
\hline \noalign {\smallskip}
Compact component	& 05$^h$ 35$^m$ 18$^s$.0525$\pm$0.0019$''$ & -05$^{\circ}$ 00$'$ 17$''$.980$\pm$0.0027$''$ & $0{\arcsec}11{{\pm}0{\arcsec}.013}{\times}0{\arcsec}.065{{\pm}0{\arcsec}.012}$, 156${\pm}$12 & 9.4$\pm$0.25 & 12 $\pm$ 0.51 \\
Extended component	& 05$h$ 35$^m$ 18$^s$.0537$\pm$0.0400$''$ & -05$^{\circ}$ 00$'$ 18$''$.167$\pm$0.048$''$ & $1{\arcsec}.9{{\pm}0{\arcsec}.12}{\times}1{\arcsec}.3{{\pm}0{\arcsec}.085}$, 150${\pm}7.5$   & 1.1$\pm$0.073 & 96 $\pm$ 6.2 \\
\hline \noalign {\smallskip}
\end{tabular}
\end{center}}
\end{table*}

\subsection{CO and SiO Line Profiles} \label{subsec:profile}
\begin{figure*}[ht!]
\epsscale{1.0}
\plotone{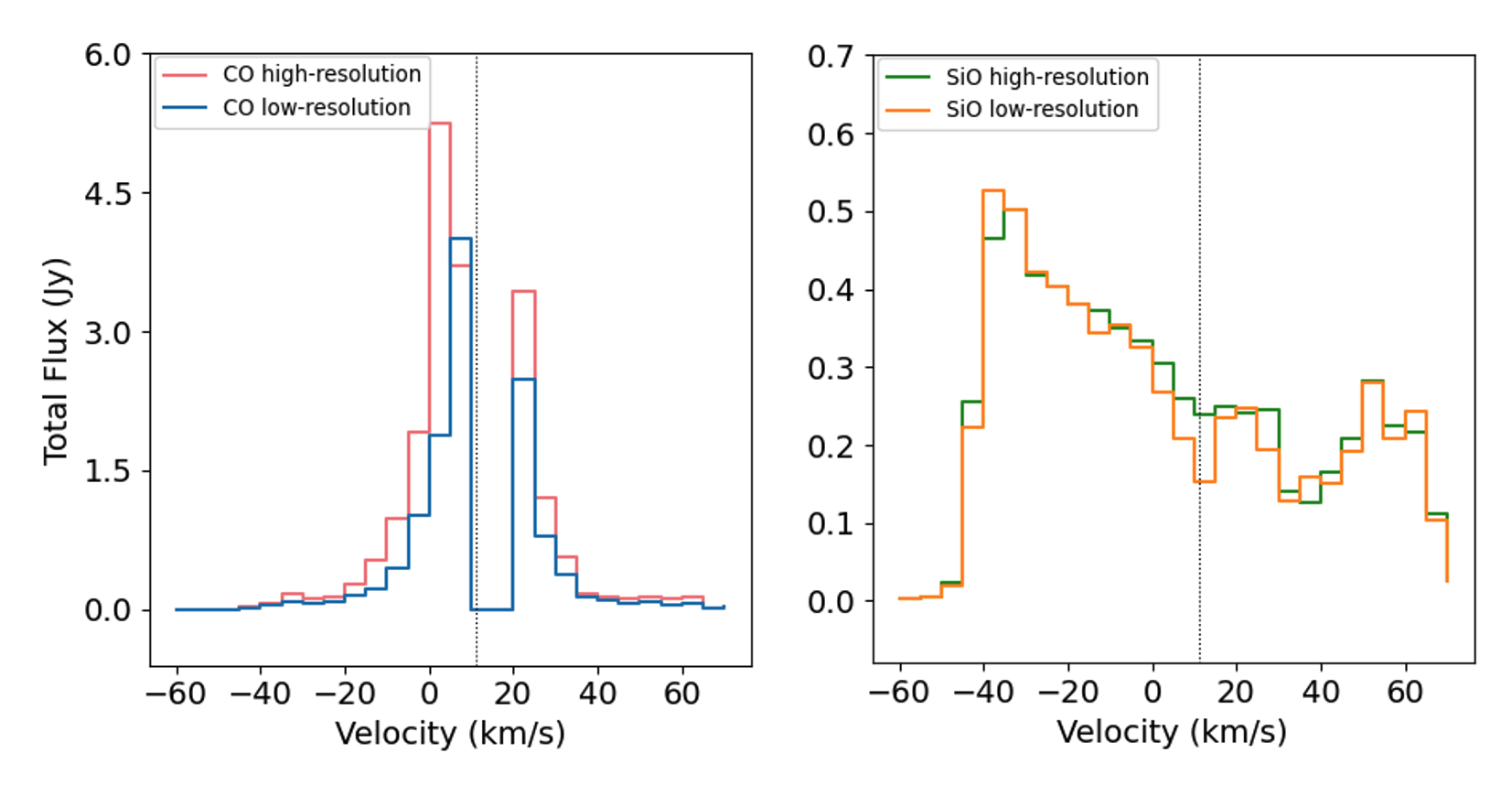}
\caption{Line profiles of MMS 1 obtained from the CO\,(2--1) 
and SiO\,(5--4) emission, respectively. 
Comparisons of the fluxes measured 
from the low- and high-angular resolution images are presented. 
The high resolution images are 
convolved with the low resolution beam size using the CASA task ``\texttt{imsmooth}''. 
Then, the images are re-gridded with ``\texttt{imregrid}''. 
Finally, flux measurements are performed for each channel 
using the area having greater than 5$\sigma$ detection 
in the high angular resolution images. 
The systemic velocity of 11\,km\,s$^{-1}$ is noted 
in the vertical dotted line in each panel.}
\label{fig:f2}
\end{figure*}

Figure~\ref{fig:f2} presents comparisons of the missing 
flux measured in the CO and SiO images obtained from 
the low resolution (${\theta}{\sim}1{\arcsec}.5$) and 
high resolution (${\theta}{\sim}0{\arcsec}.2$) images. 
The comparisons show that the fluxes obtained from the
two data sets are almost the same. 
Thus, there is no significant missing flux 
in the high angular resolution data set.  
The missing flux comparisons imply that a majority 
of the flux originating from the outflow and the jet, 
traced by CO\,(2--1) and SiO\,(5--4), came from 
compact structures that are packed within the 
MRS of the high-angular resolution data (${\sim}2{\arcsec}.0$). 
Since no significant missing flux was reported 
particularly, in the mid- and high-velocity 
ranges for both CO and SiO emissions, 
we only use the high angular resolution 
images in this paper. 

We note discrete flux enhancements both in CO\,(2--1) 
and SiO\,(5--4) at the high velocity range of 
$|v_{\rm{LSR}}-v_{\rm{sys}}|{\approx}50$\,km\,s$^{-1}$. 
These line profiles show clear evidence of 
the extremely high velocity (EHV) flow which was 
first discovered by \cite {bachiller1990} toward 
NGC\,1333 (HH\,7-11), and then was reported for several 
low- and intermediate-mass Class 0 sources 
\citep{bachiller1991a,bachiller1991b,
bachiller2000,zapata2005,hirano2010,gomez2013,gomez2019,
tafalla2017,lee2017,matsushita2019,lee2020}. 
Together with recent studies of the compact EHV flow toward MMS\,5 by \cite {matsushita2019}, 
and MMS\,6 \cite{takahashi2012}, \cite{takahashi2019}, and Takahashi et al. (2024 in prep.), the EHV flow associated with MMS\,1 is one of the most compact EHV flows ever reported (see Section~\ref{subsec:outflowjet}). 
Note that, hereafter, the terminology of ``jet (or SiO jet)'' will be used to refer to the EHV flow.

\subsection{Outflow and Jet} \label{subsec:outflowjet}

\begin{figure*}[ht!]
\epsscale{1.0}
\plotone{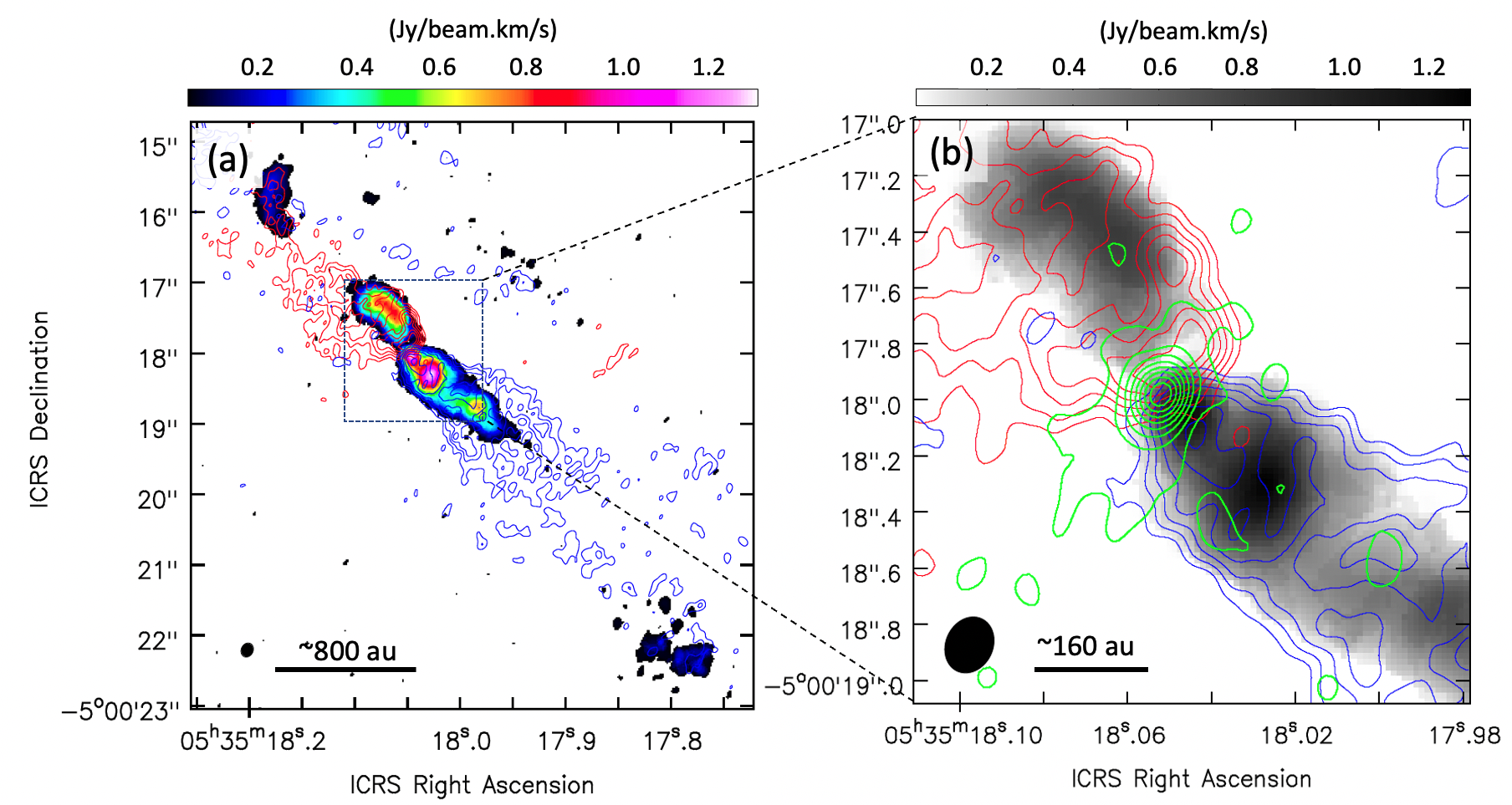}
\caption{(a) Integrated intensity image obtained from the CO\,(2--1) 
red- and blue-shifted components (red- and blue-contours) 
with the $v_{\rm LSR}$ velocity integrated over 
-45 to 10\,km\,s$^{-1}$ and 15 to 60\,km\,s$^{-1}$, respectively. 
Contour levels are 20\%, 30\%, 40\%, 50\%, 60\%, and 70\% 
with respect to the peak intensity of the blue- and red-shifted CO integrated images. 
An integrated intensity image obtained from the SiO\,(5--4) is shown in color with the $v_{\rm LSR}$ velocity integrated over 
-60 to 70\,km\,s$^{-1}$. 
(b) Zoomed in image of panel (a). 
The 1.3\,mm continuum emission is presented in green contours.
Contour level starts from 5$\sigma$ with an interval of 5$\sigma$ 
up to 40$\sigma$ (1$\sigma$=240\,mJy\,beam$^{-1}$). 
SiO integrated image is presented in the grey scale 
instead of the color used in panel (a).
The synthesized beam size is denoted 
in the bottom left corner of each panel.}
\label{fig:f3}
\end{figure*}

We present the ALMA CO\,(2--1) and SiO\,(5-4) emission 
arising from the outflow and jet located in MMS\,1 in Figure~{\ref{fig:f3}}.  
A compact molecular outflow is traced by 
the CO emission, and a collimated high-velocity 
jet within the outflow is traced by the SiO emission. 
Both outflow and jet associated with MMS\,1 were, 
for the first time, detected thanks to the ALMA's sensitivity 
and high-angular resolution capabilities. 
This clearly indicates that MMS\,1 
contains a protostar. 
The blue- and red-shifted gas are located 
in the southwest (SW) and northeast (NE) directions 
with respect to the millimeter source peak, respectively, 
with the position angle of -137$^{\circ}$. 
As seen in Figure~\ref{fig:f3}b, 
the axis of the CO outflow and SiO jet 
is aligned more or less perpendicular 
to the disk-like structure traced 
in the 1.3\,mm continuum emission. 
As clearly seen in Figure~{\ref{fig:f3}}a, 
the jet traced in the SiO emission 
shows a wiggled structure. 

\begin{figure*}
\epsscale{1.15}
\plotone{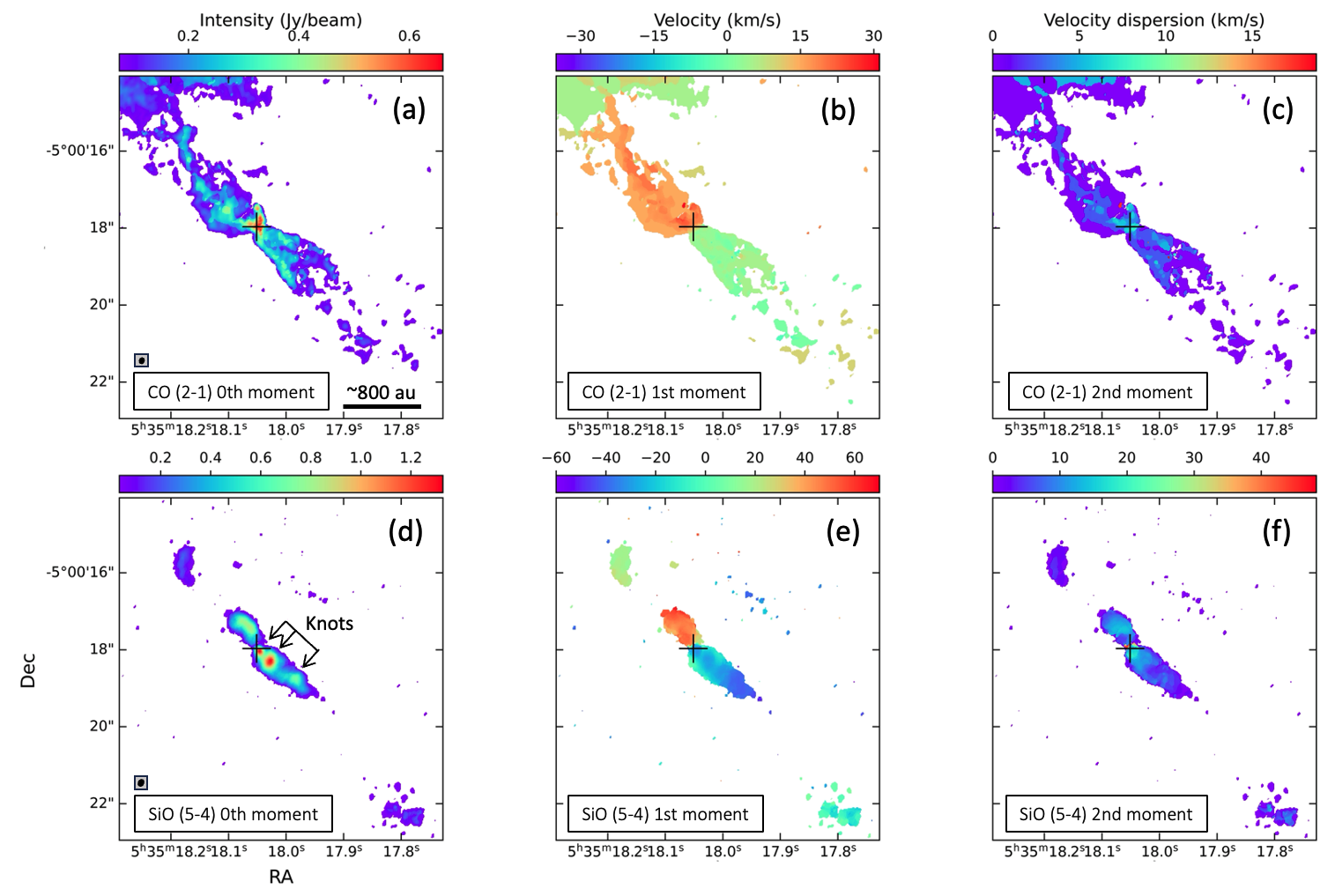}
\caption{The zeroth-, first-, and second-moment maps 
of MMS\,1 obtained from the CO\,(2--1) and SiO\,(5--4) emission. 
The moment maps were made using the CASA task \texttt{``immoments''}. 
No clip and a clip level of 0.007\,Jy\,beam$^{-1}$ 
was applied for CO\,(2--1) and SiO\,(5--4), respectively.  
Location of the 1.3\,mm continuum peak position 
is denoted by the cross mark in each panel. 
A linear size scale is denoted in the right bottom corner 
of panel (a), and the synthesized beam size of the CO and SiO 
observations are denoted in the bottom left corner 
of panels (a) and (d), respectively.}
\label{fig:f4}
\end{figure*}

Figures~\ref{fig:f4}a, b, and c present 
moment maps obtained from the CO\,(2--1) emission. 
The CO blue- and red-shifted lobes are 
extended up to ${\sim}3{\arcsec}.2$ ($\sim$1260\,au) 
and ${\sim}5{\arcsec}.1$ ($\sim$2000\,au), respectively. 
The CO emission is detected with the LSR velocity range between 
-35 and 60\,km\,s$^{-1}$.
Figure~{\ref{fig:f4}}c shows that the majority of the gas 
detected in CO\,(2--1) emission has a velocity dispersion of 
$|v_{\rm{LSR}}-v_{\rm{sys}}|{\lesssim}7$\,km\,s$^{-1}$. 
An X-shaped structure associated with central protostar, 
tracing the outflow cavity is clearly seen. 
The width of the outflow, where the outflow has the widest 
width in both lobes, is measured to be ${\sim}1{\arcsec}.3$ ($\sim$510\,au).

As presented in Figures~\ref{fig:f4}d, e, and f, 
the collimated jet is detected in SiO\,(5--4) emission 
with the projected length of 
${\sim}2{\arcsec}.3$ (${\sim}900$\,au; blue-shifted emission) and 
${\sim}1{\arcsec}.3$ (${\sim}510$\,au; red-shifted emission), respectively. 
The SiO emission was detected with 
the LSR velocity between -60 and 70\,km\,s$^{-1}$.
Velocity tendency is consistent with 
that observed in CO, while the SiO emission mainly tracing 
the high-velocity collimated jet rather 
than the wide-opening angle outflow seen in CO. 
Note that the SiO emission is located close to 
the central star, in particular 
the blue-shifted component shows a width 
comparable to the CO outflow 
(Figure~\ref{fig:f3}b and \ref{fig:f4}e).   
This wide width SiO emission distribution is only seen 
in the relatively low-velocity 
(the LSR velocity between -15\,km\,s$^{-1}$ and 5\,km\,s$^{-1}$), 
tracing the outflow cavity. 
Within the outflow cavity, we see a high-velocity collimated jet.  

\begin{figure*}
\epsscale{1.1}
\plotone{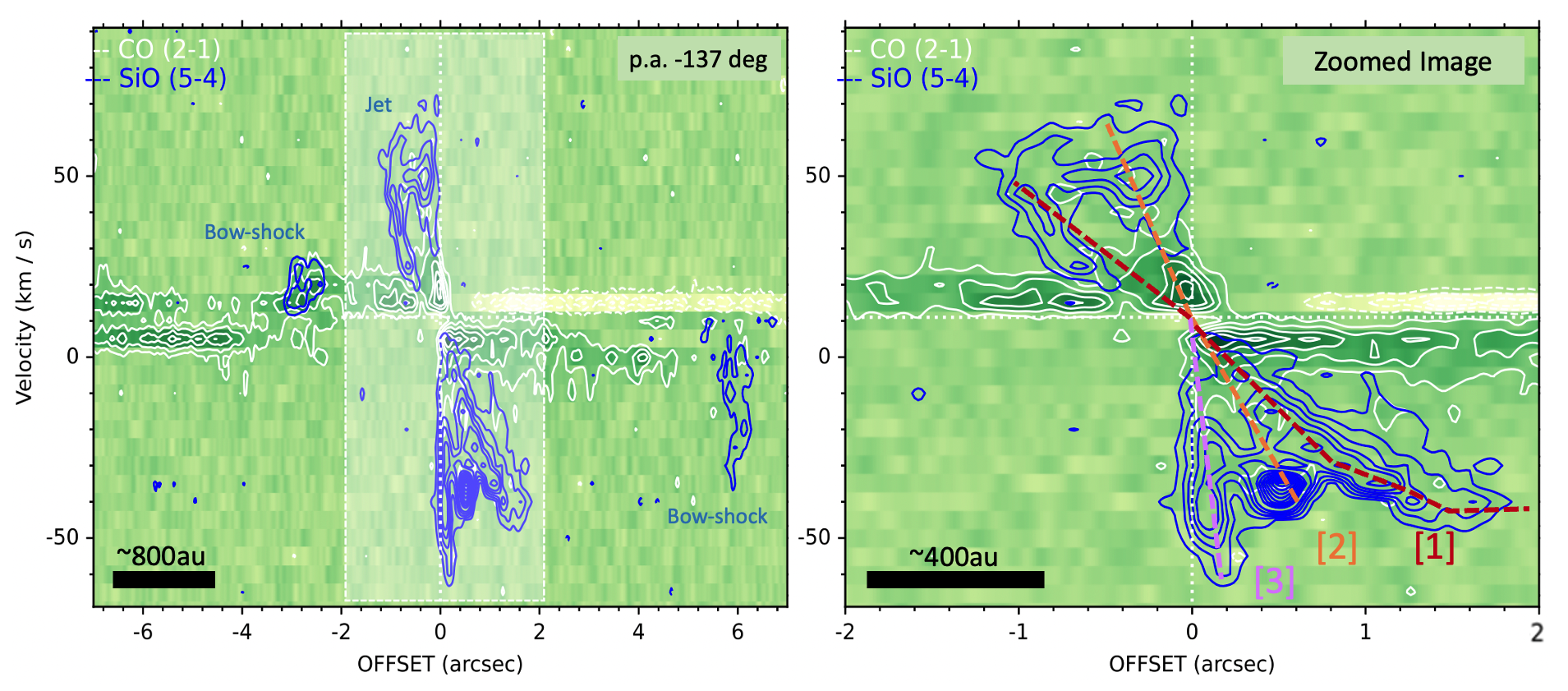}
\caption{Position-Velocity (PV) diagram cutting 
along the axis of the outflow (p.a.~=~$-137^{\circ}$).  
The background color and white contours 
correspond to CO\,(2--1) emission. 
Contour levels starts from -9$\sigma$ 
with an interval of 3$\sigma$ up to 15$\sigma$. 
The SiO\,(5--4) emission is denoted in blue contours. 
Contour levels starts from -3$\sigma$ 
with an interval of 3$\sigma$ up to 33$\sigma$. 
Negative contours are denoted by the dashed lines. 
The left panel presents the entire PV diagram, 
while the zoomed in image within the white 
hatched area is presented in the right panel, 
which reveals the emission mainly from the protostellar jet. 
The vertical and horizontal doted lines 
correspond to the location of the 1.3\,mm 
continuum emission peak and the system 
velocity of MMS\,1 ($v_{\rm sys}$=11\,km\,s$^{-1}$), respectively. 
Colored dashed lines denoted in the right panel 
are guide lines used for explaining the possible episodic mass 
ejection scenario discussed in Section~\ref{subsec:mhdsim}. 
The length of the black solid line in the bottom left corner in each panel shows a linear scale, and the width of the line shows the velocity 
resolution of the presented SiO and CO data.
}
\label{fig:f5}
\end{figure*}

The detected high-velocity SiO emission traces two components: 
(i) emission associated with the central region, 
showing a collimated structure and 
(ii) bullet-shaped emission (i.e., 
bow-shocks as explained using Figure~\ref{fig:f5} in the following paragraph) 
located at the tips of the CO outflow. 
The first component traces a collimated jet 
and is slightly brighter on the SW side (blue-shifted emission) 
than on the NE side (red-shifted emission). 
The spatially resolved image (Figure~\ref{fig:f4}d) reveals 
that the jet appears to wiggle, 
and that the SW side of the jet contains at least 
three bright knots (and an additional faint knot 
as presented in Figure~\ref{fig:f6}). 
The second component, showing the bullet-shaped structure, 
is located at ${\sim}6{\arcsec}.2$ ($\sim$580\,au; blue-shifted emission) 
and ${\sim}2{\arcsec}.0$ ($\sim$790\,au; red-shifted emission) 
away from the 1.3\,mm continuum peak position 
and brighter in the NE side (red-shifted emission) 
than in the SW side (blue-shifted emission). 
The location of the SiO bullet-shaped structure 
roughly coincides with the tips 
of the outflow lobes observed in the CO emission. 
The associated gas velocity is lower than that observed 
in the jet. 
The blue-shifted emission is detected 
in the LSR velocity range between -10 and -40\,km\,s$^{-1}$, 
and the red-shifted emission is detected 
in the LSR velocity range between 10 and 30\,km\,s$^{-1}$ 
(Figure~\ref{fig:f4} and \ref{fig:f5}). 
Velocity dispersions of ${\Delta}v{\sim}$10\,km\,s$^{-1}$ 
and ${\Delta}v{\sim}$4\,km\,s$^{-1}$ are observed 
toward the blue-shifted and the red-shifted bullet-shaped emission, 
respectively (Figure~\ref{fig:f4}f).  

Figure~\ref{fig:f5} presents the position-velocity diagram 
(PV-diagram) obtained from the CO\,(2–1) and SiO\,(5–4) data cubes 
cut along the outflow axis (p.a.~=~-137$^{\circ}$).
Clearly seen in the PV-diagram, there are  two distinct components: 
(i) a collimated jet and (ii) the bullets located at the tips 
of the CO outflow. 
Regarding the first component 
(i), SiO emission is concentrated in the central $2{\arcsec}$ region, 
showing a linear velocity increase with respect to 
distance from the center.  
The right panel of Figure~\ref{fig:f5} presents 
the zoomed image of the jet, which clearly 
shows that there are at least three 
components (as denoted by colored lines) 
showing similar respective velocity increases. 
This is particularly clear in the blue-shifted jet (i.e., SW side). 
The terminal point of each component coincides with 
an emission peak, which also corresponds to 
individual knots indicated in Figure~\ref{fig:f4}d. 
Regarding the second component (ii), 
the SiO emission is located at the tips of 
the CO blue- and red-shifted outflows. 
The spatial and velocity patterns in the PV-diagram 
are consistent with those explained 
with the jet bow-shock model presented 
in Figure~2 of \cite{arce2007}. 
Consistent with what we see 
in Figure~\ref{fig:f4}b and c, 
the CO emission in Figure~\ref{fig:f5} 
mainly traces the low- to intermediate-velocity gas components 
($|v_{\rm LSR}-v_{\rm sys}|{\lesssim}$20\,km\,s$^{-1}$). 
These components are spatially extended and mostly 
associated with the outflow cavity and lobes. 
The CO emission is also detected toward the bow-shocks, 
and is particularly bright in 
the red-shifted emission component (i.e., NE side).

\subsection{Timescales}
In this subsection, we estimate two types of timescales: 
(i) outflow and jet dynamical timescales, which represent 
how young the protostar is, and (ii) dynamical timescales 
to each knot, which is possibly related to 
the time interval of the episodic mass ejection 
within the jet. 

\subsubsection{Outflow and Jet Dynamical Timescales} \label{subsubsec:dynamicaltime}
Two velocity components are seen 
in the PV-diagram mainly obtained 
from the SiO\,(5--4) emission presented in Figure~\ref{fig:f5}. 
The first one is the SiO emission associated with 
a collimated jet, 
and the second one is the emission located 
at the tips of the outflow lobes, tracing bow-shocks. 
The dynamical timescale of the outflow and 
jet, $t_{\rm{dyn}}$, are estimated using the distance to the bow-shocks and 
length of the SiO jet, respectively. 
Gas velocities measured in each location were used for 
the calculations as following,  
\begin{equation}
t_{\rm{dyn}} = \left(\frac{l_{\rm obs}}{v_{\rm obs}}\right)\left(\frac{\cos i}{\sin i}\right),
\end{equation}
where $l_{\rm obs}$ and $v_{\rm obs}$ are the projected jet 
and outflow lengths and the line of sight gas velocities, and
$i$ is the inclination angle of the jet and outflow 
with respect to the line of sight. 
Here, we adopt $i=45^{\circ}$, estimated from 
the axis ratio of the dust disk by \cite{Liu2023}.
Given the SiO jet lengths ($l_{\rm obs}$) of 
900 au (blue-shifted) and 510 au (red-shifted side) 
measured from the first-moment map presented in Figure~{\ref{fig:f4}}d, 
and the maximum gas velocities ($v_{\rm obs}$) at the tip of the jet 
of 65~km\,s$^{-1}$(blue-shifted emission) and 
44~km\,s$^{-1}$ (red-shifted emission) 
measured from the first-moment map presented in Figure~\ref{fig:f4}e, 
the jet dynamical timescale is estimated 
to be 66~yr (blue-shifted jet) and 
55~yr (red-shifted jet), respectively. 
For the molecular outflow, we assume that 
the outflow emission is extended up to 
the locations of the bow-shocks, 
which correspond to the extension of the CO outflow. 
Given the outflow lengths ($l_{\rm obs}$) of 
2460\,au (blue-shifted side) and 1260\,au (red-shifted side), 
and the maximum gas velocities ($v_{\rm obs}$) 
measured from the SiO bow-shocks, 
presented in Figure~\ref{fig:f4}e, 
of 33\,km\,s$^{-1}$ (blue-shifted emission) 
and 9\,km\,s$^{-1}$ (red-shifted emission), 
the outflow dynamical timescales are estimated 
to be 350~yr (blue-shifted side) 
and 660~yr (red-shifted side), respectively. 
Note that the inclination angle, $i$, 
of the disk-like structure derived from 
the 1.3 mm continuum emission is $\sim$53$^{\circ}$. 
Adopting $i=53^{\circ}$ makes the change in jet/outflow lengths 
and velocities of a factor of $\sim$1.1 smaller and $\sim$1.2 larger 
with respective to the current assumptions, respectively. 
This introduces the timescale estimations of 
a factor of $\sim$1.4 longer than the current estimations.

The short dynamical timescales of the outflow 
and jet are reported by \cite{matsushita2019} torward MMS\,5/ OMC-3, which are 1300~yr and 110~yr, respectively. 
The dynamical timescales estimated for 
MMS\,1 are even shorter than those values. 
In addition, a similar compact outflow has been reported toward MMS\,6/OMC-3, which has the dynamical timescale of $\sim$50\,yr \citep{takahashi2012b}. The dynamical timescale of the SiO jet detected in MMS\,1 is as short as that of the outflow detected toward MMS\,6.

\subsubsection{Dynamical Timescale to Each SiO Knot} \label{subsec:jetknots}

\begin{figure}[thb]
\epsscale{1.1}
\plotone{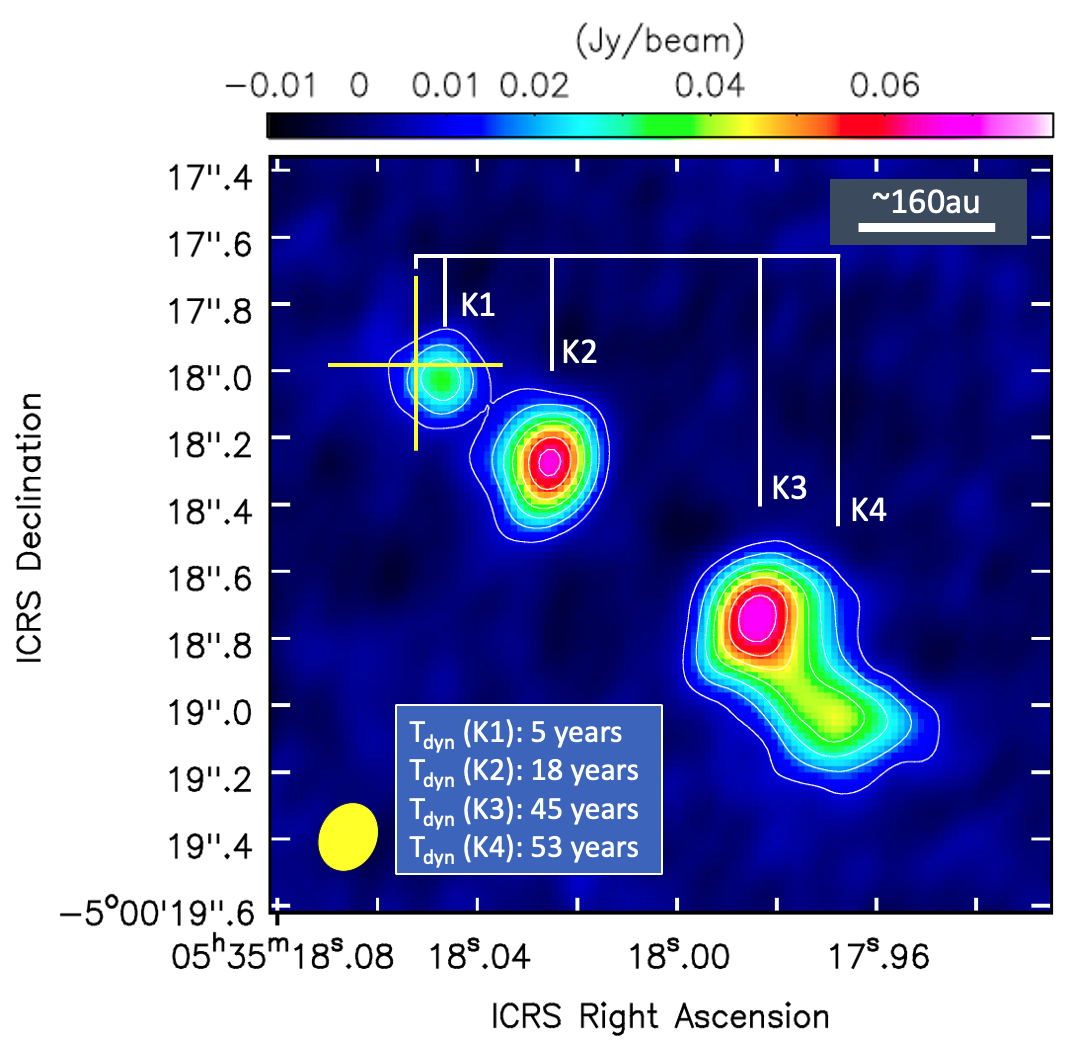}
\caption{Four bright knots, K1, K2, K3, and K4, 
detected in SiO\,(5--4) emission within the jet. 
A representative velocity channel map of 
$v_{\rm LSR}$=-40\,km\,s$^{-1}$ is presented here. 
Peak position of the millimeter emission is 
denoted by a yellow cross and dynamical timescales 
for each knot calculated in 
Subsection~\ref{subsec:jetknots} are 
listed at the bottom left corner together with the synthesized beam size. Contour levels start from 5$\sigma$ with an interval of 5$\sigma$ up to 35$\sigma$ (1$\sigma$=1.8\,mJy\,beam$^{-1}$).} 
\label{fig:f6}
\end{figure}

As presented in Figure~{\ref{fig:f4}}d and discussed 
further below in Section~\ref{subsec:mhdsim}, 
we have detected periodic flux enhancements within the SiO jet, 
which are considered to be associated with knots and possibly related to 
episodic mass ejection events. 
Figure~\ref{fig:f6} presents the map of a representative 
velocity channel ($v_{\rm LSR}{\approx}$-40\,km\,s$^{-1}$), 
showing four bright knots detected within the blue-shifted SiO jet. 
Dynamical timescales toward each knot are calculated 
using the distance between the central source (i.e., millimeter peak position) and the flux peak of each knot, and the maximum gas velocity associated with each component measured 
from the PV-diagram (Figure~\ref{fig:f5}), which are 5~yr (K1), 18~yr (K2), 45~yr (K3), and 53~yr (K4) 
after applying the inclination correction (i.e., $i$=45$^{\circ}$ estimated by \citealt{Liu2023}).

\subsection{Jet Precession} \label{subsec:jetkprec}

We showed evidence for jet wiggling in Section \ref{subsec:outflowjet} (Figure \ref{fig:f3}a). In general, jet wiggling can be explained either by orbital motion of a binary system \citep{Masciadri2002} or by jet precession (or precession of the jet-driving object; \citealt{Frank2014}). 
Both models produce a ``garden-hose'' effect, resulting in jet wiggling \citep{Raga1993}. The observed MMS\,1 jet morphology allows us to distinguish the origin of the jet wiggling \citep[][and references therein]{Frank2014,Schutzer2022}.
In the case of  jet wiggling due to binary orbital motion, the jet is expected to bend in the direction opposite to the location of the secondary (binary) component.
This produces a V-shaped jet in the vicinity of the protostar.
The previously ejected gas propagates within the cone-shaped structure, and the side-to-side wiggling 
over time produces a W-shaped structure in the projected jet locus.
Alternatively, in the case of jet precession, the jet is expected to be ejected symmetrically. Side-to-side moving gas within the cone therefore produces an S-shaped locus over time.

The SiO jet ejected from MMS\,1 shows an S-shaped wiggle. Furthermore, at the current angular resolution of ${\sim}0''.15$ ($\sim$60\,au in the linear size scale), we did not find any evidence that MMS\,1 is a binary system. These facts support that the jet wiggling associated with MMS\,1 can be caused by the jet precession rather than the orbital motion of a binary system. 

We fitted the wiggling jet observed in MMS\,1 with a three dimensional spiral morphology using the following parametric equations (3), (4), and (5),  as described by\cite{Schutzer2022} and originally based on \cite{Raga1993} and \cite{Masciadri2002}; 
\begin{equation}
x = z \tan{\beta} \sin\left({\frac{2\pi}{\lambda}}|z|+\phi\right),
\end{equation}
\begin{equation}
y = z \tan{\beta} \cos\left({\frac{2\pi}{\lambda}}|z|+\phi\right),
\end{equation}
where we consider the situation that the jet precesses inside a cone of main axis $z$. Here, $\beta$ is the jet half-opening angle, $\lambda$ is the spatial period, and $\phi$ is the phase angle. 
The precession axis tilts at an inclination angle of $i$ with respect to the plane of the sky. The projected coordinates (${\alpha}'$, ${\delta}'$) in this plane are given by 
\begin{equation}
({\alpha}', {\delta}') = (y, z \cos{i} - x \sin{i}).
\end{equation}

\begin{figure}[htb]
\epsscale{1.1}
\plotone{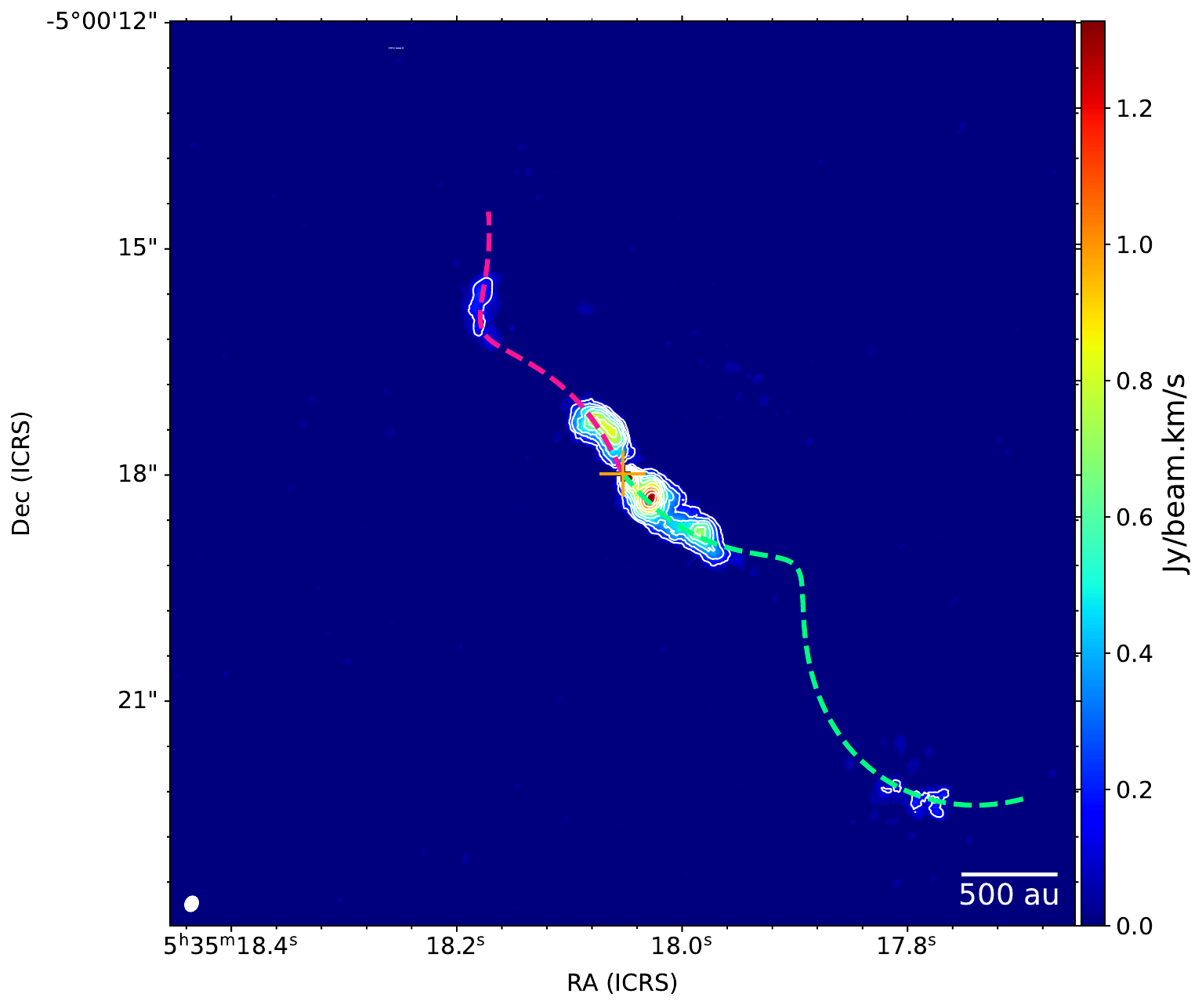}
\caption{Integrated intensity image obtained from the SiO (5--4) emission overlaid with the jet precession model discussion in Section \ref{subsec:jetkprec}. Adopted models for the northern and southern jets are denoted in magenta and light green color, respectively. Contour level starts with  10\% with an interval of 10\% up to 90\% with respect to the SiO peak intensity. Location of the 1.3\,mm dust continuum peak is marked in the orange cross. Synthesized beam is denoted in the bottom left corner.}
\label{fig:f7}
\end{figure}

We adopted the inclination angle of the jet as 45$^{\circ}$ which is estimated from the disk axis ratio, as discussed in Section \ref{subsubsec:dynamicaltime}. 
Then, the other parameters of the jet position angle $\lambda$, $\beta$, and the phase angle $\phi$ were fitted simultaneously by eye. During the fitting process, we found that the northern and southern jets have different position angles of 35$^{\circ}$ and 51$^{\circ}$, respectively, requiring an axis misalignment of 16$^\circ$. Hence, the fitting was done independently for the northern and southern jets. The fitting results and parameters are presented in Figure \ref{fig:f7} and Table \ref{table:jetparms}, respectively. Although the axis is misaligned between the northern and southern jets, the fitted parameters are similar, indicating that the jet shows a more or less symmetric ejection.

Although the cause of the jet precession is not fully understood, a few possible scenarios have been proposed. The first one is the tidal interaction between the associated protostellar disk and the noncoplanar binary companion, creating an S-shaped wiggle jet in the vicinity of protostars \citep{Terquem1999}. 
However, we did not find any apparent substructures within the central 100\,au scale with the current angular resolution, as shown in Figure \ref{fig:f1}b.  
Thus, we can exclude the tidal interaction scenario. 

Alternatively, recent simulation studies have  demonstrated that the misalignment of the core rotation axis with respect to the global magnetic field can produce an S-shaped wiggling jet  \citep{Hirano2019,Hirano2020,Machida2020}. 
In their core-collapse simulations, the precession of the protostar and disk occurs when the prestellar core has a rotation axis that is not parallel with the global magnetic field. 
In our observations, a slight misalignment between the large-scale magnetic field (p.a. $\sim$45$^{\circ}$) and both the disk rotation axis (i.e., perpendicular to the disk major axis, p.a. $\sim$65$^{\circ}$) and the observed SiO jet axis (p.a. 35$^{\circ}$ for the northern jet and p.a. 51$^{\circ}$ for the southern jet) has been observed toward MMS\,1, which could explain the jet precession observed in MMS\,1. 

\begin{table*}[htb]
{
\begin{center}
\caption{\small Parameters of jet precession model }
\label{table:jetparms}
\begin{tabular}{lccccc}
\hline \hline \noalign {\smallskip}
Parameter & $\lambda$ & $v_{\rm{jet}}$ & $\beta$ & $t_{\rm{prec.}}$ & $\phi$ \\
          &       ($''$)     & (km\,s$^{-1}$)  &   ($^{\circ}$) &  (yr) & ($^{\circ}$) \\
\hline
North & 7.8  &   62  &   7.0 &   234 & 174  \\
South & 6.8  &   92  &   8.0 &   138 & 180  \\
\hline \noalign {\smallskip}
\end{tabular}
\end{center}}
\footnotesize {\bf{Notes.}}  
The jet velocity was measured from the SiO observations. The values measured from the first-moment map (see Section \ref{subsubsec:dynamicaltime}) were used to calculate the jet velocities with  $v_{\rm{jet}}$=$v_{\rm{obs}}$/$\sin{i}$ (with $i$=45$^{\circ}$).\\
\end{table*}

\vspace{1cm}

\section{Discussion} \label{sec:discussion}

The molecular outflow and jet associated with MMS\,1 
are detected for the first time with the high-angular 
resolution and the high-sensitivity capabilities of ALMA. 
The detected jet is the smallest thus far reported, hence 
a good opportunity to look into the initial mass ejection phenomena 
from a protostar. 
As presented in Figure~\ref{fig:f5}, 
the jet velocities linearly increase 
as the distance from the center increases.
In addition, Figure~\ref{fig:f6} indicates 
the possibility of episodic mass ejection events.
In this section, we discuss possible origins of the velocity structure by comparing the observational results with MHD simulations (Subsection~\ref{subsec:mhdsim}). 
Then, we discuss how the presented dynamical model relates to 
those features traced by the SiO emission (Subsection~\ref{subsec:shockchem}). 
Finally, we discuss the intermittent mass ejection phenomena 
and how this may be related to 
episodic mass accretion, which has previously been 
reported for several protostars 
(Subsection~\ref{subsec:episodicevents}).  

\subsection{Comparison between Observation and Simulation} \label{subsec:mhdsim}
\begin{figure*}
\epsscale{1.1}
\plotone{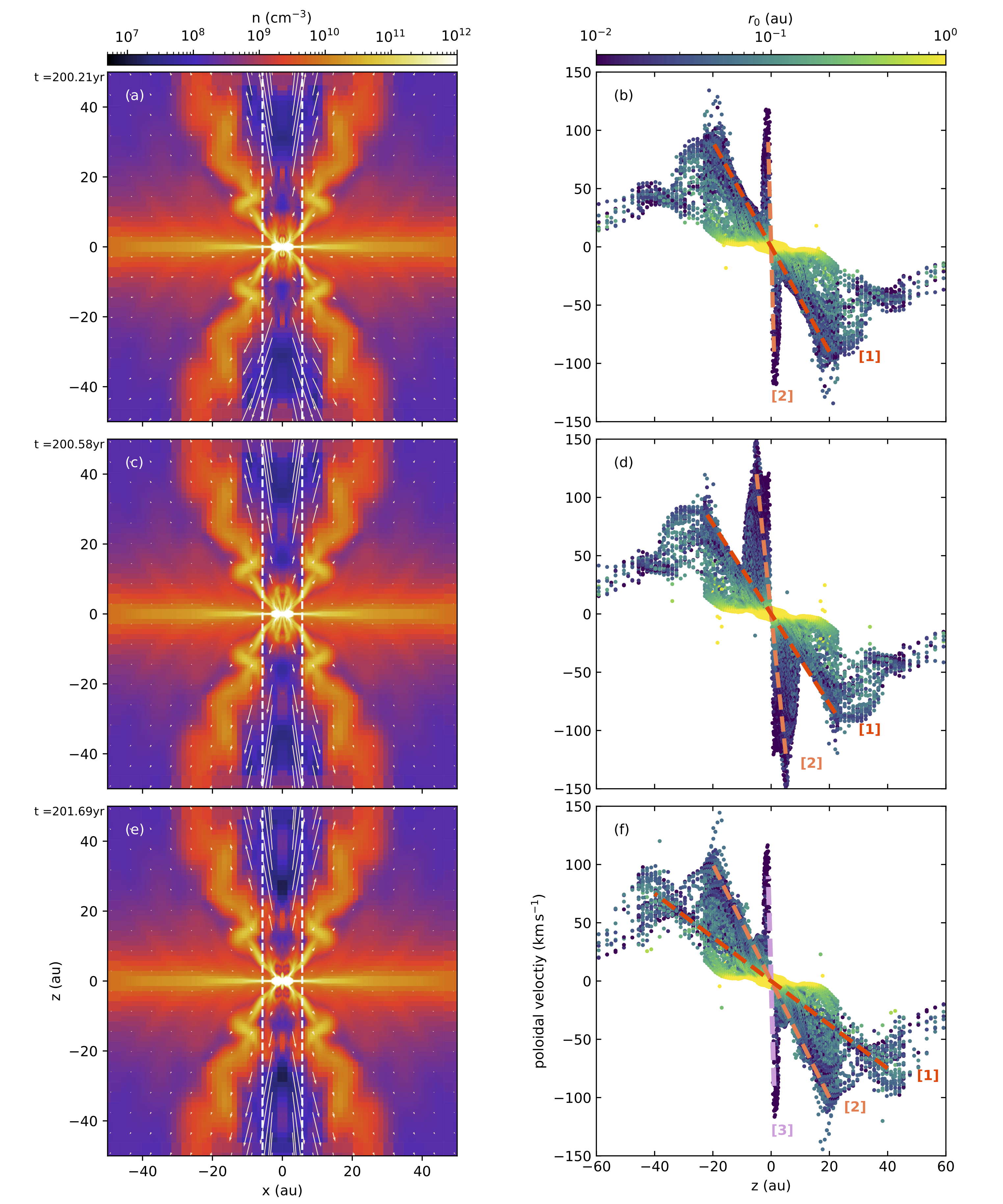}
\caption{
Time evolution of the density (color) and 
velocity (arrows) distributions (left panels) 
and the PV-diagram along the $z-$axis (right panels) made from the numerical simulations done by \citet{machida2019}.
The elapsed time after protostar formation $t$ is described at the upper left corner of the left panels.
In the right panels, the color within each circle corresponds to 
the jet launching radius calculated by the physical quantities derived from the simulation and analytical solutions in \citet{anderson2003}.
The spine structures corresponding to the ejected clumps are labeled as [1], [2] and [3] in the right panels. 
For comparison with observational results (corresponding to labels [1], [2], and [3] in Figure~\ref{fig:f5}), three recent  ejection events are delineated by the orange dashed lines in the PV diagram of panel (f).   
}
\label{fig:f8}
\end{figure*}

To further validate the acceleration of the jet in MMS\,1 seen in Figure \ref{fig:f5},
we compare a recent numerical simulation of protostar evolution with our results.
The data in Figure~\ref{fig:f8} were taken from the core collapse simulation
performed by \citet{machida2019}.

We first explain the numerical settings and detailed simulation results by \citet{machida2019}, for which a spherical core with a Bonnor-Ebert density profile was adopted as the initial state. 
The radius $R_{\rm cl}$ and mass $M_{\rm cl}$ of this initial cloud core are $R_{\rm cl}=1.2\times10^4$\,au and  $M_{\rm cl}=1.7\,M_\odot$.
A uniform magnetic field with $B_0=4.5\times10^{-5}$\,G and a rigid rotation of $\Omega_0=1.1\times10^{-13}$\,rad\,s$^{-1}$ were added to the initial state.
The gravitational collapse of the core was calculated using a nested grid method \citep{Machida2004, Machida2013}, in which the resistive magnetohydrodynamics equations (see, eqs. [1]--[7] in \citealt{Machida2012}) were solved. 
The finest spatial resolution of the simulation is $5.6\times10^{-3}$\,au. 
As a result, both the star-forming core ($\sim10^4$\,au) and the protostar ($\sim0.01$\,au) were spatially resolved in the simulation. 
Starting from the prestellar core, the evolution of the system is calculated for 2000\,yr after protostar formation. 

The simulation produces episodic jets driven from the inner disk region \citep[see Fig.~2 of ][]{machida2014}. 
These jets with speeds $\sim100$\,km\,s$^{-1}$ appear repeatedly every $\sim1-10$\,yr until the end of the simulation 2000\,yr after protostar formation. 
In the following analysis, we consider the simulation result between 200.21-201.69\,yr after protostar formation, during which the high-speed jet appears three times, as shown in Figure~\ref{fig:f8}.
Although the full simulation reveals a wide range of mass ejection rates due to the jets ($\sim10^{-6}-10^{-9}$\,M$_\odot$\,yr$^{-1}$, see Fig.~7 of \citealt{machida2019}), the mass ejection rate during the epoch shown in Figure~\ref{fig:f8} is about $3\times 10^{-6}$\,M$_\odot$\,yr$^{-1}$.

In Figure~\ref{fig:f8}, the density and velocity distributions on the $y=0$ plane (left) and the PV diagram along the jet or $z-$axis (right) at three different epochs, 200.21, 200.58, and 201.69\,yr after protostar formation 
are plotted. 
As described above, the simulation spatially resolved the protostar and circumstellar disk and reproduced the outflow and jet.
Thus, we can qualitatively compare the kinematics 
of the observed jet with the simulation. 
The left panels of Figure~\ref{fig:f8} show that 
the high-velocity flow, with several local density peaks, is surrounded 
by the low-velocity flow (for details, see \citealt{machida2019}).
The right panel of Figure~\ref{fig:f8} plots the PV-diagram at each epoch, in which we only used the simulation data 
around the z-axis as -5\,au ${\leq} x {\leq}$ 5\,au 
and -5\,au$ {\leq} y {\leq}$ 5\,au to create the PV-diagram 
(the white dashed line in panels (a), (c), and (e)) 
in order to focus on the high-velocity component 
(i.e., the jet).

In the second and fourth quadrants of the PV diagram in Figure~\ref{fig:f8}b, we can confirm several spine-like components.  
Each spine corresponds to a clump ejected 
from the region near the protostar. 
We can see a weak density peak at $z\simeq{\pm}20$\,au within the white dashed line of Figure~\ref{fig:f8}a which corresponds to the clump  produced during a previous mass ejection event occurring $197.18$\,years after protostar formation. 
In the PV diagram (Figure~\ref{fig:f8}b), we can also confirm 
the jet intermittent components at $z=\pm20$\,au, 
corresponding to the clump made during the previous ejection event, which is labeled as [1]. 
In addition, a strong mass ejection is currently occurring at the root of the jet. 
A sharp spine around $z\simeq0$ in the PV diagram (Figure~\ref{fig:f8}a) labeled as 
[2] corresponds to the current mass ejection event close to the protostar. 
The color in the PV diagram representing the launching radius of each velocity component indicates that the jet launching radii are widely  distributed across the range of 0.01-1\,au. 
We confirmed that various velocity components in the range $<150$\,km\,s$^{-1}$ appear at every mass ejection event. 
The mass ejection occurs at different disk radii and the gas ejected at different radii has different velocities. 
Thus, a sharp spine appears around $z=0$ in the PV diagram immediately after the mass ejection occurs. 
Note that mass ejection recurrently occurs in the main accretion phase \citep{machida2019}.

The spine labeled as [2] widens and is inclined from the vertical axis in Figure~\ref{fig:f8}(d). 
In addition, we can confirm that the jet velocity labeled as [2] linearly increases as the distance from the protostar increases in Figure~\ref{fig:f8}(f), which is similar to the Hubble-like velocity
structure or Hubble flow.
As described above, the ejected gas has different velocities and the gas with different velocities propagates along the $z-$axis. 
Thus, the highest-velocity gas is located farthest from the protostar as time proceeds.
On the other hand, the low-velocity gas remains close to the protostar because their initial velocity (i.e., launching velocity) is low.  
As a result, the inclined spine-like structure appears in the PV diagram. 
In addition, the ejected gas interacts with the envelope material and thus decelerates. 
Therefore, a sharp spine widens with time and the propagation or jet velocity decelerates to have a velocity plateau, as seen in the right panels of Figure~\ref{fig:f8}. 

In Figure~\ref{fig:f8}(f), we can see a new spine labeled as [3] in the PV diagram which corresponds to a new mass ejection event occurring at $201.69$\,yr after the protostar formation (Figure~\ref{fig:f8}e). 
The spline inclines as time in the PV diagram. 
Reflecting the past mass ejection event, we can see, at least, three spike-like structures [1]-[3] in Figure~\ref{fig:f8}(f). 
The spine gradually inclines and widens with time. 
In other words, the inclination angle and width of the spine at an old mass ejection event are more significant than that at a new mass ejection event.

Some morphological similarities between the 
observation (Figure~\ref{fig:f6}) and simulation 
(Figure~\ref{fig:f8} left panels) are apparent. 
In addition, the PV diagrams made by 
the observation (Figure~\ref{fig:f5}) 
and simulation (Figure~\ref{fig:f8}) are qualitatively the same. 
However, the velocity distribution and the interval of the mass ejection events are not quantitatively the same between the observation and the simulation. 
Since the spatial resolution of the simulation is restricted, the timescale over which the PV diagram varies is extremely short (months). We thus can not quantitatively compare the observation with the simulation.

Following the simulation, we interpret 
the observed PV diagram velocity distribution. 
Multiple mass ejection events produce the multiple spine-like structures seen in Figure~\ref{fig:f5}. 
The different angles seen in the PV diagram can be explained by the time evolution of each jet component. 
The high-velocity gas component spreads fastest as the ejected gas moves away from the protostar.
Thus, in the PV diagram the spine appears to incline with time. 
In addition, the interaction between the ejected gas and envelope material decelerates the jet and produces a velocity plateau. 
In Figure~\ref{fig:f5}, for component [1] the farthest part of the spine reaches 
a constant velocity (i.e., velocity plateau), 
potential evidence for the decelerated gas. 
Such a velocity plateau has been reported in other protostellar systems (e.g., \citealt{Wang2014,matsushita2019,Morii2021b}). 

\subsection{Jet Dynamical Model and Shock Chemistry}
\label{subsec:shockchem}

The PV diagram obtained from the simulation shows similar characteristics to those obtained from the SiO observation.
However, the 
simulation PV diagram seems to differ from that found by previous outflow studies using CO, which mainly traces an older and static outflow, such as reported by \cite{Santiago-garcia2009} and \cite{plunkett2015}. Hence, we consider that the chemical processes in this extremely young jet are very active and may not be in a steady state. 
In this subsection, we discuss possible origins of the SiO emission observed toward MMS\,1.
We also discuss how they possibly relate to the mass ejection phenomena produced by the dynamical model presented in Subsection \ref{subsec:mhdsim}. 

Three different paths are considered to release silicates into the gas phase and subsequently form the SiO molecules observed in jets \citep{Cabrit2012, Podio2021}. 
The first path is the sputtering of silicate grains in C-shocks, which requires a velocity
differential 
in the range of 10 and 40 km\,s$^{-1}$ to produce SiO in the gas phase (e.g., \citealt{Schilke1997, Caselli1997, Panoglou2012}). 
The second path is the release of silicon into the gas phase by vaporization in grain-grain collisions, which is expected to occur in slow J-shocks ($\lesssim$50\,km\,s$^{-1}$) with the amount of silicon released into the gas phase on the order of a few percent \citep{Guillet2009}. 
The third path is the release of silicon into the gas phase 
in the region where the gas temperature reaches the dust sublimation temperature \citep{Glassgold1991, Tabone2020}. 

The first case (C-shocks) has been discussed mainly for forming SiO in 
previously observed sources 
\citep{Schilke1997, Gusdorf2008b, Gusdorf2008a} for which SiO is known as a  shock tracer.
The second case (J-shocks) can be expected to occur in specific locations, such as at the apex of bow shocks (e.g., \citealt{Hartigan2004}). 
The PV-diagrams in the right panels of Figure \ref{fig:f8} indicate that the jet velocity is consistent with that required to release the silicate grains into the gas phase either through the C- or J-shocks. 
Moreover, the numerical simulation performed by \cite{Machida2020} reveals that the region close to the jet driving region has density and temperature discontinuities, indicating the presence of J-shocks.
These considerations imply that the observed SiO emission might be produced by both the first and second pathways above, C-shocks or J-shocks. 

Regarding the third pathway, the launching point of the high-speed jet ($v_{\rm{jet}}{\sim}$100\,km\,s$^{-1}$) is expected to be several $\times$ 0.1~au (Figure \ref{fig:f8} right panels). 
The temperature expected at these radii is 
a few $\times$ 1000\,K, assuming a source bolometric luminosity of 1\,L$_{\odot}$.  
This temperature is comparable to the dust sublimation temperature of 1500\,K (e.g., \citealt{Vaidya2009}), implying that some of the observed high-velocity SiO components might be formed by this additional route. 

Our gas-dynamic model 
(left panels of Figure \ref{fig:f8})
does not consider chemical networks. 
However, the physical conditions expected from this model are consistent with the conditions under which the SiO molecule could be formed, as predicted by the previous theoretical studies described above.
Chemohydrodynamic analysis, where the chemical distribution is affected by the gas-dust dynamical evolution  (e.g., \citealt{Flower2003, Godard2019, Castellanos2018}), will be required to make complete comparisons in future studies.

\subsection{Episodic Mass Ejection and Accretion} \label{subsec:episodicevents}

As discussed in Section~\ref{subsec:mhdsim}, 
both observations and simulations clearly show 
evidence of the non-steady 
mass ejection events within protostellar jets. 
Recent ALMA CO and SiO observations toward 
protostellar sources reveal chains of knots within protostellar jets, suggesting episodic mass ejection events
(\citealt{plunkett2015,matsushita2019, 
Jhan2022, Dutta2022, Dutta2023}, and this work). 
The dynamical timescales between these ejection events range from
several years to thousands of years.
At the same time, recent flux variability surveys performed 
at sub-millimeter and mid-infrared wavelengths 
suggest that the fraction of variable sources 
is high during this protostellar phase   
(\citealt{Johnstone2018, YHLee2021, Park2021}).
Two distinct forms of variability are observed: years-long secular variability  
and short-timescale stochastic variability. 
The range of timescales is  attributed to a variety of processes including underlying changes in the disk accretion rate, geometric changes in the circumstellar disk, 
hydromagnetic interactions between the stellar surface and inner disk, magnetic reconnection of the stellar magnetosphere, 
and the number or size of spots on the stellar surface \citep{Park2021, Fischer2023}. 
In conclusion, these monitoring studies demonstrate the importance of understanding the non-steady nature the star formation activities. 

\begin{figure*}[htb]
\epsscale{0.9}
\plotone{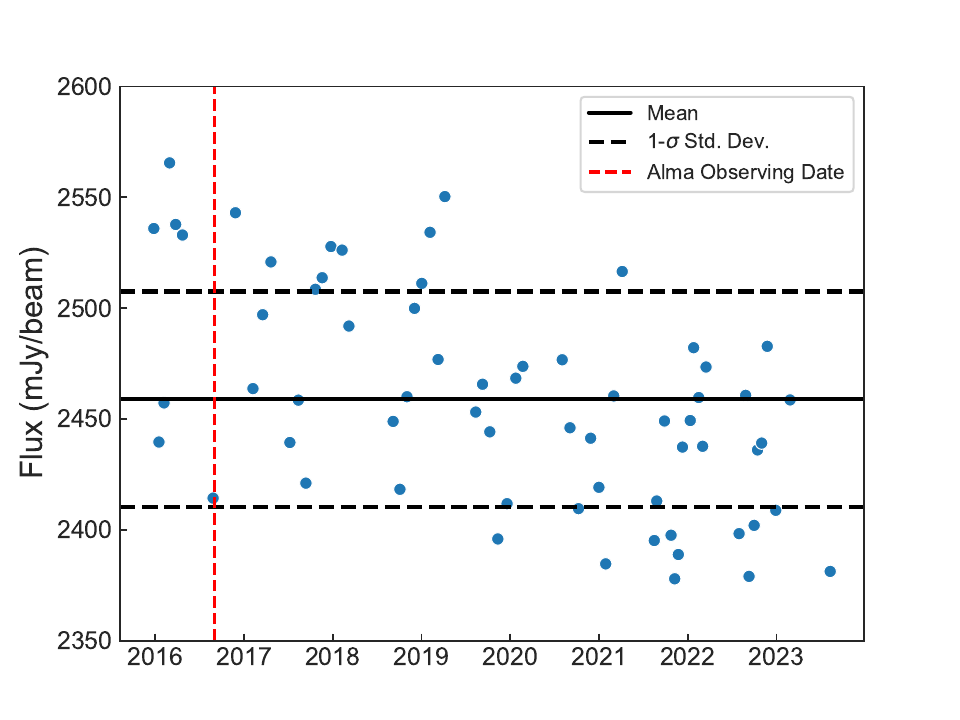}
\caption{The 850\,$\mu$m light curve obtained with the JCMT/SCUBA2. 
The horizontal dashed lines give the uncertainties of $\pm$1$\sigma$ due to calibration on individual measurements. 
The vertical red dashed line shows the date 
at which the high-angular resolution ALMA observation, 
presented in this paper was performed.}
\label{fig:f9}
\end{figure*}

Figure~\ref{fig:f9} presents 
850\,$\mu$m continuum flux monitoring results for MMS\,1  obtained by 
the JCMT Transient Survey \citep[][]{Herczeg2017}. The observations cover late 2015 through mid-2023, almost eight years of roughly monthly cadence. For bright sources, such as MMS\,1, the relative flux calibration uncertainty between epochs is better than 2\% \citep[][Mairs et al.\ 2024 accepted by ApJ]{Mairs2017}.
No drastic flux change is detected 
from the 850\,$\mu$m light curve, suggesting that there has not been a clear burst event within the last 8 years. There is, however, a distinct $\sim$ 0.5\% per year gradual decline in the brightness  - therefore a candidate linear variable, according to Mairs et al.\ (2024 accepted).  The decline is similar, though shallower, to the observed decline in HOPS\,383 \citep{YHLee2021}, a protostar that had an observed mid-IR burst between 2004 and 2006 \citep{Safron2015}, about ten years prior to the start of the JCMT survey.

For MMS\,1, the timescale of the episodic mass ejection events associated with the SiO knots presented 
in Figure~\ref{fig:f6} ranges from 5 to 50 years. Thus, the monitored time period might not be long enough to detect the associated episodic 
mass accretion phenomena. We note that 850\,$\mu$m burst events have been observed by the JCMT Transient Survey for a few protostars allowing for detailed consideration of the events. HOPS 373 underwent a months-long burst that was analysed by \citet{Yoon2022} and EC53 (also known as V371 Ser) has continuing quasi-periodic variations observed at both near-IR \citep{Hodapp2012} and sub-mm \citep{Yoo2017} leading to a multi-wavelength analysis by \citet{YHLee2020} and an interferometric analysis by \citet{Francis2022}. Thus far, however, neither of those two sources have had features in their outflow connected to the observed accretion bursts.  Circumstantial evidence for accretion variability correlating with episodic mass ejection has been presented for a few Orion outflow sources through comparison of the ejection timescales and the observed accretion variability timescale found from mid-IR or sub-mm monitoring \citep{Jhan2022, Dutta2023}.
Continued monitoring of MMS\,1 at 850\,$\mu$m, combined with proper motion 
measurements of the SiO knots, will enable us to better determine the connection between episodic mass accretion events 
and episodic mass ejection.

Considering the current ALMA angular 
resolution of ${\sim}0{\arcsec}.2$ and the observed 
maximum jet velocity of ${\sim}$70\,km\,s$^{-1}$, 
our observations are only sensitive to 
the timescale longer than $\sim$5~years. 
In order to detect less significant episodic 
events associated with a shorter time period 
than five years, higher angular resolution 
observations (${\lesssim}0{\arcsec}.2$) and 
mild shock tracers such as 
CH$_3$OH and H$_2$CO or S-bearing 
molecules might be useful 
(e.g., \citealt{Lee2018,Tychoniec2019, Tychoniec2021, Codella2020,Codella2021}).

\section{SUMMARY} \label{sec:sum}

We have performed CO\,(2--1), SiO\,(5--4), and 1.3\,mm continuum observations using ALMA Band 6 toward an extremely young intermediate-mass protostellar source, MMS\,1 located in the OMC-3 region. 

\begin{enumerate}
    \item	We have detected bright compact 1.3\,mm continuum emission. Assuming that the compact component fitted by the 2D Gaussian traces a dust disk, the size and mass of the dust disk are estimated to be ${\sim}33$\,au and 0.0051-0.020\,M$_{\odot}$, respectively. 
    
    \item	With the angular resolution of ${\sim}0{\arcsec}.2$, we have detected the very compact molecular outflow in CO\,(2–1) and the jet in SiO\,(5-4) for the first time. The outflow and jet are aligned roughly perpendicular to the dust disk. The detected molecular outflow shows a cavity like structure with a wide opening angle. The SiO jet lies within the CO cavity.
    The jet length and velocity are calculated to be $\sim$1300\,au and 92\,km\,s$^{-1}$ for the blue-shifted component, and 720\,au and 62\,km\,s$^{-1}$ for the red-shifted component after the correction for inclination. We confirm that the SiO jet wiggles. Within the jet knots are observed, which are particularly clear to the SW side of the jet.    
     
    \item   The jet dynamical time scale is estimated to be 66\,yr (blue-shifted component) and 55\,yr (red-shifted component). These numbers indicate that the jet associated with MMS\,1 is the youngest thus far reported. The dynamical timescale of each SiO knot with respect to the millimeter source peaks (i.e., protostar location) are estimated between 5\,yr and 53\,yr, possibly related to the timescale of episodic mass accretion events.  

    \item The PV-diagram cut along the SiO jet axis shows that the gas velocity linearly increases as the distance from the protostar increases (a.k.a. Hubble-like velocity structure). The SiO emission associated with each knot shows a different inclination angle (or slope) in the PV-diagram. A numerical simulation of core collapse by \cite{machida2019} can qualitatively reproduce similar features in the PV-diagram. The comparison suggests that the observed SiO jet may be explained by multiple-mass ejection events, which produce the multiple-spine-like structure observed in the PV-diagram.  
   
    \item Finally, the 850\,$\mu$m light curve obtained with the JCMT/SCUBA~2 toward MMS\,1 is presented. The plot indicates no significant time variability, such as large burst or drop, over the last eight years. A weak decrease in the flux with time might be related to changes in the mass accretion rate, similar to that observed for HOPS 383. Continued sub-mm flux monitoring combined with proper motion measurements of the jet knots may reveal how episodic mass ejection events are connected to the episodic mass accretion history of MMS\,1. 

\end{enumerate}

{\it Acknowledgements.}

We thank the anonymous referee for providing very
helpful comments and useful suggestions, which improve the manuscript significantly. This paper makes use of the following ALMA data: ADS/JAO.ALMA\#2015.1.00341.S. ALMA is a partnership of ESO (representing its member states), NSF (USA) and NINS (Japan), together with NRC (Canada), MOST and ASIAA (Taiwan), and KASI (Republic of Korea), in cooperation with the Republic of Chile. The Joint ALMA Observatory is operated by ESO, AUI/NRAO and NAOJ. Data analysis was carried out on the Multi-wavelength Data Analysis System operated by the Astronomy Data Center (ADC), National Astronomical Observatory of Japan. 
We acknowledge the JCMT transient team providing us the eight years light curve obtained in the 850\,$\mu$m continuum emission using SCUBA\,2, which was used to compare with our ALMA data set in Figure~\ref{fig:f9}. 
This work was supported by JSPS KAKENHI Grant Number JP19K03919(KT), and JP21K03617 and JP21H0046 (MNM). L.A.Z. acknowledges financial support from CONACyT-280775 and UNAM-PAPIIT IN110618, and IN112323 grants, M\'exico. D.J.\ is supported by NRC Canada and by an NSERC Discovery Grant.

{\it Facilities:ALMA; JCMT} 

\vspace{5mm}




\end{document}